\documentclass[format=acmsmall, screen=true, review=false]{acmart}
\usepackage{lineno,hyperref}
\usepackage{booktabs} 
\usepackage{multirow}
\usepackage{graphicx}
\usepackage{longtable}
\usepackage{todonotes}
\usepackage{pdflscape}
\usepackage{comment}
\usepackage{colortbl}
\usepackage{hhline}
\modulolinenumbers[5]
\usepackage{caption}
\usepackage{subcaption} 

\copyrightyear{2023}
\acmYear{2023}
\setcopyright{rightsretained}
\acmDOI{10.1145/3585004}

\usepackage{multibib}
\newcites{S}{Mapping study references}
\usepackage{tcolorbox}
\tcbuselibrary{many}
\usepackage{xcolor, soul}
\sethlcolor{lightgray}  
\tcbset{
    enhanced,
    breakable,
    attach boxed title to top left={
        xshift=0.5cm,
        yshift=-\tcboxedtitleheight/2},  
    top=4mm,
    coltitle=white,
    beforeafter skip=\baselineskip,
}
\newenvironment{mybox}[1]{%
    \begin{tcolorbox}[title={Key Future Questions~--~#1}]%
    }{
    \end{tcolorbox}
    }
\begin{document}

\title{Modern Code Reviews - A Survey of Literature and Practice}
\author{Deepika Badampudi}
\email{deepika.badampudi@bth.se}

\author{Michael Unterkalmsteiner}
\email{michael.unterkalmsteiner@bth.se}

\author{Ricardo Britto}
\email{ricardo.britto@bth.se}
\affiliation{
    \institution{Blekinge Institute of Technology}
    \department{Software Engineering Research and Education Lab Sweden}
    \streetaddress{Valahallavägen 1}
    \city{Karlskrona}
    \postcode{37141}
    \country{Sweden}
}
\additionalaffiliation{
    \institution{Ericsson AB}
    \streetaddress{Ölandsgatan 1}
    \city{Karlskrona}
    \postcode{37133}
    \country{Sweden}
}

\begin{abstract}
\textbf{Background:}  Modern Code Review (MCR) is a lightweight alternative to traditional code inspections. While secondary studies on MCR exist; it is unknown whether the research community has targeted themes that practitioners consider important.
\\
\textbf{Objectives:} The objectives are to provide an overview of MCR research, analyze the practitioners' opinions on the importance of MCR research, investigate the alignment between research and practice, and propose future MCR research avenues.
\\
\textbf{Method:} We conducted a systematic mapping study to survey state-of-the-art until and including 2021, employed the Q-Methodology to analyze the practitioners' perception of the relevance of MCR research, and analyzed the primary studies' research impact.
\\
\textbf{Results:} We analyzed 244 primary studies, resulting in five themes. As a result of the 1300 survey data points, we found that the respondents are positive about research investigating the impact of MCR on product quality and MCR process properties. In contrast, they are negative about human factor- and support systems-related research.
\\
\textbf{Conclusion:} These results indicate a misalignment between the state-of-the-art and the themes deemed important by most survey respondents. Researchers should focus on solutions that can improve the state of MCR practice. We provide an MCR research agenda, which can potentially increase the impact of MCR research.
\end{abstract}



\maketitle

\section{Introduction}\label{sec:introduction}

Software code review is the practice that involves the inspection of code before its integration into the code base and deployment. Software code reviews have evolved from being rigorous, co-located, and synchronous to lightweight, distributed, tool-based and asynchronous~\cite{Sadowski:2018}. Modern Code Review (MCR) is a lightweight alternative to traditional code inspections~\cite{fagan1976}, which focuses on code changes and allows software developers to improve code quality and reduce post-delivery defects~\cite{5, Bacchelli2013}. MCR is an essential practice in modern software development not only due to its contribution to quality assurance; it also helps with design improvement, knowledge sharing, and code ownership~\cite{nazir2020modern}.

The research interest on code inspections declined in the middle of the 2000's~\cite{kollanus2009survey}. Due to the value of code reviews in general, it is reasonable to assume that the research focus has shifted to MCR. After over a decade of research on MCR, several initiatives were born to aggregate a body of knowledge on the increasing research of this essential quality assurance practice. To the best of our knowledge, we presented in our previous work~\cite{badampudi_modern_2019} the first overview on the state-of-art of MCR research. In our previous mapping study, we reported the preliminary results of systematically searching and analyzing the existing literature (based on titles and abstracts) and identified major research themes. Likely in parallel, other studies have also explored and made an attempt to aggregate the existing literature on MCR, either on particular aspects of the practice (refactoring-aware code reviews~\cite{coelho2019refactoring}, benefits of MCR~\cite{nazir2020modern}, MCR in education~\cite{indriasari2020review}, reviewer recommendations~\cite{ccetin2021review}) or in general~\cite{wang2021can, davila2021systematic}.

Since there exists a considerable and diverse amount of research on the MCR practice, we were curious whether the research community has targeted themes that are also perceived as important by MCR practitioners. Similar investigations have been conducted in the past on software engineering research in general~\cite{lo2015practitioners, carver2016practitioners} and requirements engineering research in particular~\cite{franch2020practitioners}.

The main goal of this study is therefore \emph{to provide an overview of the different research themes on MCR, analyze practitioners' opinions on the importance of the research themes, and outline a roadmap for future research on MCR}. To achieve this goal, we extended our earlier work~\cite{badampudi_modern_2019} by including publications up until the year 2021 and synthesizing the contributions of the 244 identified primary studies in MCR research. Then we constructed 47 statements that describe the research covered in the primary studies and surveyed 28 practitioners using the Q-Methodology~\cite{zabala2016bootstrapping} to gauge their perception on the statements representing the research conducted in this field. Finally, we compare the practitioners perception on the investigated themes in MCR research with the amount of publications and research impact of those themes. The main contributions of this paper are as follows:
\begin{itemize}
    \item \textbf{A comprehensive aggregation of research conducted on MCR research themes until and including 2021}~--~We identify potential gaps that researchers could address in the future and provide a summary on the state-of-the-art in MCR research that can be useful for practitioners (e.g., to benefit from existing findings and solutions).
    \item \textbf{Level of alignment between MCR state-of-the-art and practitioners' perception on the relevance of the MCR state-of-the-art }~--~We assess the practitioners' perception on the relevance of the MCR state-of-the-art represented by statements that summarize each topic in the MCR state-of-the-art. We assess the alignment between what the research community has focused on the most and how MCR practitioners perceive its relevance. This analysis can help researchers to focus on themes that are deemed relevant by practitioners but do not have enough research coverage. We propose a research roadmap based mainly on the analysis of the reviewed primary studies, and qualified by the responses from the survey. 
\end{itemize}
The remainder of this paper is structured as follows: Section~\ref{sec:related work} presents background on the MCR practice and relevant related work to this study. Section~\ref{sec:RM process} describes the design of our research, which is followed by Sections~\ref{Sec:msresults} and~\ref{sec:surveyresults}, where we describe the mapping study and survey results, respectively. In Section~\ref{sec:comp}, we compare the state-of-the-art and practitioners' perspectives. Section~\ref{sec:discussion} discusses our results and illustrates our MCR research roadmap. Finally, Section~\ref{sec:conclusions} presents our conclusions and view on future work.

\section{Background and Related work}\label{sec:related work}
\begin{figure*}[h]
    \centering
    \includegraphics[width=\textwidth]{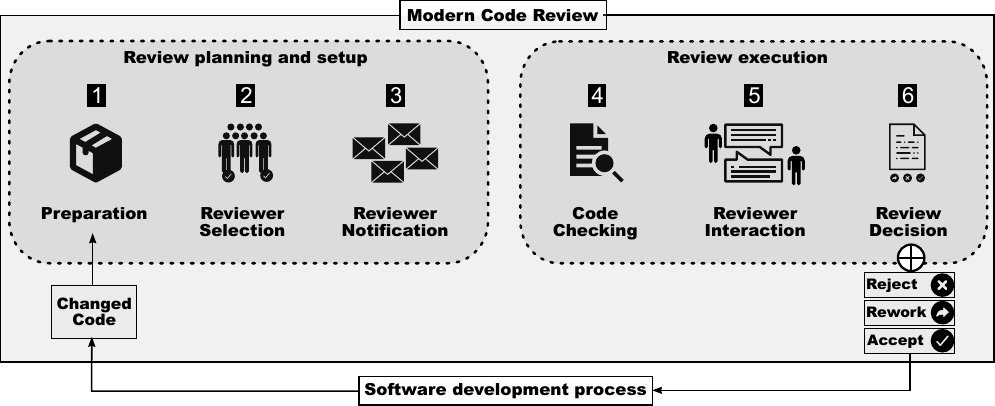}
    \caption{Overview of steps in modern code reviews (adapted from Davila and Nunes~\cite{davila2021systematic})}
    \label{fig:mcr}
\end{figure*}

In this section, we briefly revise the history of peer code reviews (\ref{sec:pcr}), illustrate the MCR process (\ref{sec:background}), discuss related work surveying MCR literature (\ref{sec:literature}) and practitioner surveys in SE in general (\ref{sec:practice}), illustrating the research gap that this study aims to fill (summarized in~\ref{sec:gap}).

\subsection{Peer Code Review}\label{sec:pcr}
It is widely recognized that the peer review of code is an effective quality assurance strategy~\cite{aurum2002state, kollanus2009survey}. Formal code inspections were introduced in the mid-1970s by Fagan~\cite{fagan1976}. The formal code inspection process requires well-defined inputs (a software product that is ready for review), planning and organizing of review resources (time, location, expertise), execution of the review following guidelines that facilitate the detection of defects, and a synthesis of the findings, which are used for improvement~\cite{aurum2002state}. Kollanus and Koskinen~\cite{kollanus2009survey} have reviewed the research on code inspections between 1980 and 2008 and found a peak of publications between the late 1990s and 2004 (averaging 14 papers per year) with a strong decline between 2005 and 2008 (4 papers per year). This change in research interest coincides with the rise of MCR research, starting around 2007, which had a steady upward trend since 2011~\cite{badampudi_modern_2019}. Research on code inspections focused on reading techniques, effectiveness factors, processes, the impact of inspections, defect estimation, and inspection tools~\cite{kollanus2009survey}. Interestingly, the tool aspect was the least researched one, with 16 out of 153 studies (10\%). Modern code reviews (MCR) were born out of the need to perform lightweight yet efficient and effective quality assurance~\cite{Bacchelli2013}. It is a technology-driven practice that complements continuous integration and deployment (CI/CD), a method to frequently and reliably release new features. CI/CD also saw a rise in practical adoption and research interest around 2010~\cite{shahin2017continuous}. 

\subsection{Modern Code Review}\label{sec:background}
Figure~\ref{fig:mcr} illustrates the two phases and main six steps in MCR, which are typically supported by tools that integrate with version control systems (e.g., Gerrit, GitHub, and GitLab). The main actors involved in MCR are the code author(s) and the reviewer(s). While there may be organizational, technical and tooling differences between open source and commercial software development implementing MCR, the general steps are valid for both contexts. A significant difference of MCR in open source and commercial development is its perceived purpose. In open source development, reviewers focus on building relationships with core developers, while in commercial development, knowledge dissemination through MCR is more important~\citeS{bosu2016process}.

In \emph{Step 1}, the code author(s) prepare the code change for review, which usually includes a description of the intended modification and/or a reference to the corresponding issue recorded in a bug tracking system. When tools like Gerrit and  GitHub are used, the change author creates a pull request. 
Questions that arise in this step are: What is the optimal size of a pull request? How can large changes be broken down into manageable pull requests? How should changes be documented? 

In \emph{Step 2}, the project/code owner selects one or more reviewers, typically using heuristics such as expertise in the affected code or time availability. Questions that arise in this step are: What is the optimal number of reviewers? Who is the best reviewer for a particular code change? What is the optimal workload for a reviewer? How should code changes be prioritized for review?

In \emph{Step 3}, the reviewer(s) are notified of their assignment, concluding the planning phase of MCR. Questions that arise in this step are: How much time should be allocated to a review? How should reviews be scheduled (batch, scattered throughout the work day/week)?   

In \emph{Step 4}, the reviewer(s) check the code changes for defects or suggest improvements. The procedure for this step is highly individualized and depends on tool support, company culture, and personal preference and experience. Typically, the reviewer inspects the code changes that are surrounded by unchanged (context) code and provide feedback to particular lines of code, as well as to the overall change. Questions that arise in this step are: What information is needed and what are the best practices for an effective review? What is the most effective way to describe findings and comments to code changes? Can the identification of certain defects or improvements be automated? 

In \emph{Step 5}, the reviewer(s) and author(s) discuss the code changes and feedback, often facilitated by tools that enable asynchronous communication and allow referencing code and actors. This interaction creates a permanent record of the technical considerations regarding the change that emerged during the review. Questions that arise in this step are: What are the key considerations for effective communication between reviewer(a) and author(s)? How can endless (unprofessional) discussions be avoided? How can consensus be facilitated?  

In \emph{Step 6}, the change is either rejected, accepted, or sent back to the author(s) for refinement. The decision process can be implemented with majority voting or rests upon the judgement of the project/code owner. Questions that arise in this step are: To what extent can the decision process be automated? What can we learn from accepted/rejected changes that can be used to accept/filter high/low quality patches earlier?

The questions above, together with other questions, are investigated in the primary studies identified in our study, but also in the literature surveys presented in the related work, which is discussed next. 

\subsection{Literature surveys}\label{sec:literature}





While surveys on software inspections in general~\cite{laitenberger2000encompassing, aurum2002state, kollanus2009survey}, checklists~\cite{brykczynski1999survey}, and tool support~\cite{macdonald1995review} have been conducted in the past, surveys on MCR have only recently received an increased interest from the research community (since 2019). We identified six studies, besides our own, that mentioned MCR in their review aim within a very short time frame (2019-2021). Table~\ref{tab:reviewcomparison} summarizes key data of these reviews. 

\begin{table}
    \caption{Comparison of review studies on MCR}\label{tab:reviewcomparison}
    \centering
    \scriptsize
    \begin{tabular}{llllp{6.5cm}}
        \toprule
        Review by & Time-frame & Studies & Focus & Research questions \\
        \midrule
        Badampudi et al. & \multirow{2}{*}{2007-2018} & \multirow{2}{*}{177} & \multirow{2}{*}{MCR in general} & RQ1. What topics of modern code reviews are investigated?  \\
        \cite{badampudi_modern_2019} & & & & RQ2. How were the aspects in R1 investigated?\\
        \midrule
        Coelho et al. & \multirow{3}{*}{2007-2018} & \multirow{3}{*}{13} & Refactoring-aware & RQ1. What are the most common research topics?\\
        \cite{coelho2019refactoring} & & & code review & RQ2. What are the methods/techniques/tools proposed?\\
        & & & & RQ3. What are the validation methods applied?\\
        \midrule
        Nazir et al. & \multirow{2}{*}{2013-2019} & \multirow{2}{*}{51} & \multirow{2}{*}{Benefits of MCR} & RQ1. What are the real benefits of performing MCR? \\
        \cite{nazir2020modern}& & & & RQ2. How identified benefits can be grouped into relevant themes? \\
        \midrule
        Indriasari et al. & \multirow{3}{*}{2013-2019} & \multirow{3}{*}{51} & MCR as a & RQ1. What are the reported motivations for conducting peer code review activities in tertiary-level programming courses?\\
        \cite{indriasari2020review}& & & teaching vehicle & RQ2. How has peer code review been practiced in tertiary-level programming courses? \\
        & & & & RQ3. What are the main benefits and barriers to the implementation of peer code review in tertiary-level programming courses?\\
        \midrule
        \multirow{3}{*}{\c{C}etin et al.} & \multirow{6}{*}{2009-2020} & \multirow{6}{*}{29} & & RQ1. What kind of methods/algorithms are used in CRR studies?\\
        & & & Code reviewer & RQ2. What are the characteristics of the datasets in CRR studies?\\
        & & & recommendations & RQ3. What are the characteristics of the evaluation setups used in CRR studies?\\
        \cite{ccetin2021review} & & & (CCR) & RQ4. Are the models proposed in CRR studies reproducible?\\
        & & & & RQ5. What kind of validity threats are discussed in CRR studies?\\
        & & & & RQ6. What kind of future works are discussed in CRR studies?\\
        \midrule
        Wang et al. & \multirow{3}{*}{2011-2019} & \multirow{3}{*}{112} & \multirow{3}{*}{Benchmarking MCR} & RQ1. What contributions and methodologies does code review (CR) research target?\\
        \cite{wang2021can} & & & & RQ2. How much CR research has the potential for replicability?\\
        & & & & RQ3. What metric and topics are used with CR studies?\\
        \midrule
        Davila and Nunes & \multirow{3}{*}{1998-2019} & \multirow{3}{*}{138} & \multirow{3}{*}{MCR in general} & RQ1. What foundational body of knowledge has been built
based on studies of MCR?\\
        \cite{davila2021systematic} & & & & RQ2. What approaches have been developed to support MCR?\\
        & & & & RQ3. How have MCR approaches been evaluated and what were the reached conclusions?\\
        \midrule
        \multirow{3}{*}{This study} & \multirow{3}{*}{2007-2021} & \multirow{3}{*}{244} & \multirow{3}{*}{MCR in general} & RQ1. Which MCR themes have been investigated by the research community?\\
        & & & & RQ2. How do practitioners perceive the importance of the identified MCR research themes? \\
        & & & & RQ3. To what degree are researchers and practitioners aligned on the goals of MCR research?\\
        \bottomrule
    \end{tabular}
\end{table}

To the best of our knowledge, our systematic mapping study~\cite{badampudi_modern_2019} presented the first results on the state-of-the-art in MCR research (April 2019). We identified and classified 177 research papers covering the time frame between 2007 and 2018. The goal of this mapping study was to identify the main themes of MCR research by analyzing the papers' abstract. We observed an increasing trend of publications from 2011, with the major themes related to MCR processes, reviewer characteristics and selection, tool support, source code characteristics and review comments. In this paper, we update the search to include studies published including 2021 and we considerably deepen the classification and analysis of the themes covered in MCR research, reporting on the major contributions, key takeaways and research gaps. Furthermore, we survey practitioners opinions on MCR research in order to juxtapose research trends with the perspective from the state-of-practice.

Briefly after our mapping study, Coelho et al.~\cite{coelho2019refactoring} published their mapping study on refactoring-aware code reviews (May 2019). They argue that MCR can be conducted more efficiently if reviewers are aware of the type of changes and focus therefore their search on methods/techniques/tools that support the classification of code changes. They identified 13 primary studies (2007-2018), of which 9 are unique to their review. This could be due to the inclusion of "code inspection" in their search string, resulting in papers that are not related to MCR (e.g. \cite{chen2018clone,alves2017refactoring}), even though Coelho et al. mentioned MCR explicitly in their mapping aim. 

Nazir et al.~\cite{nazir2020modern} published preliminary results of a systematic literature review on the benefits of MCR in January 2020. They identified 51 primary studies, published between 2013-2019, and synthesized nine clusters of studies that describe benefits of MCR: software quality improvement, knowledge exchange, code improvement, team benefits, individual benefits, ensuring documentation, risk minimization, distributed work benefits, and artifact knowledge dissemination.

Indriasari et al.~\cite{indriasari2020review} reviewed the literature on the benefits and barriers of MCR as a teaching and learning vehicle in higher education (September 2020). They identified 51 primary studies, published between 2013 and 2019, and found that skill development, learning support, product quality improvement, administrative process effectiveness and the social implications are the main drivers for introducing peer code reviews in education. Analyzing the set of primary studies they included, we observe that this review has the least overlap of all with the other reviews. This is likely due to the particular focus on peer code reviews in education, which was explicitly excluded, for example, in our study.

\c{C}etin et al.~\cite{ccetin2021review} focused in their systematic literature review on the aspect of reviewer recommendations in MCR (April 2021). They identified 29 primary studies, published between 2009 and 2020, and report that the most common approaches are based on heuristics and machine learning, are evaluated on open source projects but still suffer from reproducibility problems, and are threatened by model generalizatibility and data integrity.

We discuss now the two reviews of MCR that are closest in scope and aim to our work and illustrate similarities in observations and the main differences in contributions between the reviews. Wang et al. published a pre-print~\cite{wang2019evolution} on the evolution of code review research (November 2019), which has been extended, peer-reviewed, and published in 2021~\cite{wang2021can}. They identified 112 primary studies, published between 2011 and 2019. Similar to our results (see Figure~\ref{fig:research_type}), they observe a predominant focus on evaluation and validation research, with fewer studies reporting experiences and solution proposals for MCR. The unique contributions of their review are the assessment of the studies' replicability (judged by availability of public data sets) and the identification and classification of metrics used in MCR research. The former is important as it allows other researchers to conduct replication studies and the latter helps researchers to design studies whose results can be benchmarked. Compared to Wang et al., our review of the themes studied in MCR research is more granular (9 vs. 47) and we provide a narrative summary of the papers' contributions.   

Finally, Davila and Nunes~\cite{davila2021systematic} performed a systematic literature review with the aim to provide a structured overview and analysis of the research done on MCR (2021). They identified 138\footnote{When we analysed their primary studies, we found they had one duplicated paper in their set, which included 139 studies.}  primary studies published between 1998 and 2019, and provide an in-depth analysis of the literature, classifying the field into foundational studies (which try to understand the practice), proposal studies (improve the practice), and evaluation studies (measure and compare practices).  
Their synthesis provides excellent insights in the MCR state-of-art with findings that are interesting for researchers as well as practitioners.

\subsection{Practitioner surveys}\label{sec:practice}
\begin{table}
    \caption{Comparison of studies analyzing practitioners' perception of research}\label{tab:surveycomparison}
    \centering
    \footnotesize
    \begin{tabular}{lll}
        \toprule
         Survey by & Source of research & Practitioner perception\\
         \midrule
         Lo et al.~\cite{lo2015practitioners} & General Software Engineering (ICSE and ESEC/FSE) & 71\% positive \\
         Carver et al.~\cite{carver2016practitioners} & Empirical Software Engineering (ESEM) & 67\% positive \\
         Franch et al.~\cite{franch2020practitioners} & Requirements Engineering & 70\% positive \\
         \bottomrule
    \end{tabular}
\end{table}

Several studies have investigated the practical relevance of software engineering by surveying practitioners (see Table~\ref{tab:surveycomparison}). Lo et al.~\cite{lo2015practitioners} were interested to gauge the relevance of research ideas presented at ICSE and ESEC/FSE, two premier conferences in Software Engineering. They summarized the key contributions of 571 papers and let practitioners (employees at Microsoft) rate the ideas on a scale from essential, worthwhile, unimportant to unwise. Overall, 71\% of the ratings were positive. However, they found no correlation between academic impact (citation count) and relevance score. Carver et al.~\cite{carver2016practitioners} replicated Lo et al.'s study with research contributions from a different conference (ESEM), targeting a wider and international audience of practitioners. They also investigated what practitioners think SE research should focus on. Their conclusions are similar to the ICSE and ESEC/FSE study, with 67\% overall positive ratings for the research, and no correlation between academic impact and relevance score. Furthermore, they found that the research at the conference addresses the needs expressed by the practitioners quite well. However, they highlight the need for improving the discoverability of the research to enable knowledge transfer between research and practice. 

Finally, Franch et al.~\cite{franch2020practitioners} surveyed practitioners in the field of requirements engineering and found mostly positive ratings (70\%) for the research in this area. The practitioners justifications for positive ratings are related to the perceived problem relevance and solution utility described in research. The requirements engineering activities that should receive the most attention from research, according to the practitioners needs, are traceability, evaluation, and automation.

\subsection{Research gap}\label{sec:gap}
While recent years have seen several literature surveys on MCR, we know very little about how this research is perceived by practitioners. Looking at the research questions shown in Table~\ref{tab:reviewcomparison}, one can observe that a few studies sought to investigate how MCR techniques~\cite{ccetin2021review} and approaches~\cite{wang2021can, davila2021systematic} have been evaluated. None has yet studied practitioner perception of MCR research, even though general software and requirements engineering research has been the target of such surveys (see Table~\ref{tab:surveycomparison}). In this study, we focus the literature review on identifying the main themes and contributions of MCR research, summarizing that material in an accessible way in the form of evidence briefings, and gauge practitioners perceptions of this research using a survey. Based on the results of this two data collection strategies, we outline the most promising research avenues, from the perspective of their potential impact on practice.  

\section{Research Design}\label{sec:RM process}
Based on our main goal introduced in Section~\ref{sec:introduction}, we formulated the following research questions.

\begin{description}
    \item [RQ1] Which MCR themes have been investigated by the research community?
    \begin{description}
        \item [RQ1.1] How was the research on MCR conducted and in which context?
        \item [RQ1.2] What is the quality of the conducted research?
        \item [RQ1.3] Which were the most investigated MCR themes and what were the major findings of the MCR research? 
    \end{description}
    \item [RQ2] How do practitioners perceive the importance of the identified MCR research themes?
    \item [RQ3] To what degree are researchers and practitioners aligned on the goals of MCR research?
\end{description} 

\begin{figure}
    \centering
    \includegraphics[width=.5\linewidth]{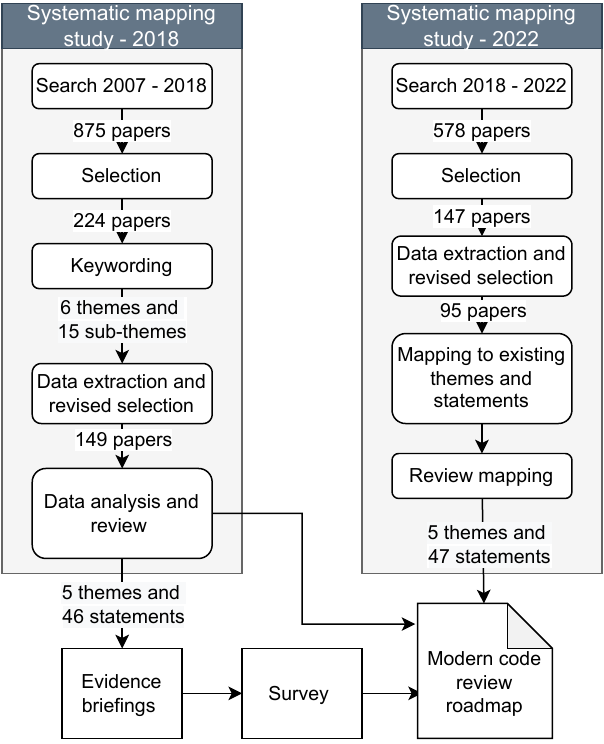}
    \caption{Research methodology followed in this study.}
    \label{fig:RM}
\end{figure}

To answer these questions, we followed a mixed-methods approach. We conducted a systematic mapping study to answer RQ1 and its sub-questions. In our previous study~\cite{badampudi_modern_2019}, we presented preliminary results of the mapping study with the review period until 2018. To answer RQ2, we created statements representing the primary studies' research objectives. We then created the survey questionnaire using the statements representing the primary studies until 2018. We conducted the survey using the Q-Methodology, collecting practitioners' opinions on the importance of the statements representing the MCR research topics. We extended the mapping study period to 2021, analyzed the new primary studies (2018 onwards), and mapped to the statements representing them. Finally, we answer RQ3 by comparing both the frequency and research impact of MCR research with its perceived importance by practitioners. The research design is depicted in Figure~\ref{fig:RM}. 
All research material we produced in this study (search results, selected primary studies, data extraction, evidence briefings, survey material, and citation analysis) is available online~\cite{deepika_badampudi_2022_7066821, deepika_badampudi_2021_briefings}. In the remainder of this section, we illustrate the research methodologies we employed to answer our research questions.

\subsection{Systematic mapping study} \label{SubSec:SMS}
We followed the guidelines for conducting systematic mapping studies by Petersen et al.~\cite{PETERSEN20151}, which include the following steps: (1) definition of review questions; (2) conduct search for primary papers; (3) screening relevant papers; (4) keywording of abstracts; (5) Data extraction and mapping of studies. As seen in Figure~\ref{fig:RM}, the systematic mapping study (SMS) was conducted in two phases: one study until the review period 2018 (SMS2018) and another from review period until 2021 (SMS2022). The review questions are represented by the sub-questions RQ1.1, RQ1.2, and RQ1.3.  Below we provide details on the execution of the consolidated mapping study (including the initial (SMS2018) and extend mapping study (SMS2022)).

\paragraph{Search} \textit{Databases:} The following databases were selected in SMS2018 and SMS2022 studies based on their coverage of papers: Scopus, IEEE Explore, and ACM Digital Library. Scopus indexes a wide range of publishers such as ScienceDirect and Springer. \\
\textit{Search Strings:} We used the keywords listed in Table~\ref{tab:Keywords} to search in the three databases, using the search strings shown in Table~\ref{tab:SearchStrings}, combining each keyword with a logical "OR" operation and adding a wildcard (*) operator. We intentionally did not include the term \emph{“inspection"} due to its association with traditional code inspections, which researchers have reviewed in the past (see discussion at the beginning of Section \ref{sec:literature}).\\ 
\textit{Search Scope:} The search results, including SMS2018 and SMS2022 studies, are presented in Table~\ref{tab:database results}. In order to consider full years results we included papers until the year 2021 in the SMS2022 study. All the search results were exported into a csv file. We identified duplicates using Microsoft Excel's conditional formatting and applying the duplicate values rule to all the titles. We manually checked the formatted duplicate entries before removing them from the list. 

\begin{table}
\caption{Search keywords and results}
\begin{subtable}[h]{0.45\textwidth}
\centering
 \scriptsize
\begin{tabular}{ll} 
\toprule
\textbf{ID} & \textbf{Keyword }                  \\
\midrule
K1 & code review               \\
K2 & patch accept              \\
K3 & commit review             \\
K4 & pull request              \\
K5 & modern code inspect       \\
\bottomrule
\end{tabular}
\caption{Keywords used in search strings}
\label{tab:Keywords}
\end{subtable}
\hfill
\begin{subtable}[h]{0.45\textwidth}
    \centering
    \scriptsize
    \begin{tabular}{ll}
    \toprule
       \textbf{Database}  & \textbf{Papers} \\
       \midrule
        Scopus & 1412 \\
        IEEE Explore & 497\\
        ACM Digital Library - Title & 104\\
        ACM Digital Library - Abstract & 291\\
        \midrule
        \textbf{Total} & 2392 \\ 
        \textbf{Total after removing duplicates and noise} & 1453 \\
        \bottomrule
    \end{tabular}
    \caption{Search results from each database}
    \label{tab:database results}
\end{subtable}
\vspace{-4mm}
\end{table}

\begin{table}
\centering
\scriptsize
\caption{Search strings used in different databases}
\label{tab:SearchStrings}
\begin{tabular}{ll} 
\toprule
\textbf{Database} & \textbf{Search string}\\
\midrule
Scopus                                                                       & TITLE-ABS-KEY ("K1*"~OR~"K2*"~OR~"K3*"~OR~"K4*"~OR~"K5*")  \\
IEEE Explore                                                                 & ("All Metadata":"K1*"~OR~"K2*"~OR~"K3*"~OR~"K4*" OR "K5*")
                             \\
\multicolumn{1}{c}{\multirow{2}{*}{ACM Digital Library - Title and abstract}} & acmdlTitle:(K1*" "K2*" "K3*" "K4*" "K5*")                \\
\multicolumn{1}{c}{}                                                          & recordAbstract:("K1*" "K2*" "K3*" "K4*" "K5*")                                \\
\bottomrule
\end{tabular}
\end{table}

\paragraph{Selection}
\begin{table}
\centering
\scriptsize
\caption{Inclusion and exclusion criteria}
\label{tab:inexcriteria}
\begin{tabular}{lll} 
\toprule
 & \textbf{ID} & \textbf{Criterion} \\
\midrule
\multirow{4}{*}{Inclusion} & I1 & Papers related to MCR in general.\\
        & I2 & Papers discussing source code (including test code) review done on a regular basis.\\
        & I3 & Papers discussing MCR related aspects (reviewer selection, benefits, outcomes, challenges, motivations, etc.) \\
        & I4 & Papers proposing solutions for particular MCR activities.\\
\midrule
\multirow{5}{*}{Exclusion} & E1 & Papers not discussing MCR or the subject of
investigation is not MCR process.\\
        & E2 & Papers that do not discuss the implications of a solution on the
MCR process.\\
        & E3 & Papers that discuss MCR in education.\\
        & E4 & Papers not in English and without accessible full text.\\
        & E5 & Panel abstracts and proceedings summaries.\\
        & E6 & Short papers without results and symposium summarising results of other published papers. \\
\bottomrule
\end{tabular}
\vspace{-4mm}
\end{table}

The search results were reviewed based on a defined set of inclusion and exclusion criteria (Table~\ref{tab:inexcriteria}). \\
\textbf{SMS2018 Selection:} Before we started the selection process, we conducted a pilot selection on randomly selected papers from the result set. All three authors performed an independent decision on whether the paper should be included, excluded or tentatively included (we decided to be rather inclusive and exclude a paper later based on reading the full text). During the first pilot on 20 papers, we noticed a paper on test case review. We refined the inclusion criterion I2 (see Table~\ref{tab:inexcriteria}) to add test code review as well. Some papers discussed approaches to support the MCR process to make it more efficient, for example, by selecting a relevant reviewer. Therefore, we added a specific inclusion criterion related to the MCR process (I4). 
As one of the goals of our study is to understand practitioners' perception on MCR research, we decided to only include studies focusing on MCR practice (including open source). Therefore we excluded papers that discuss MCR in education (E3). We modified exclusion criterion E1 to emphasize the subject of the investigation, i.e. we only include papers where the process of MCR is under investigation. We also came across papers that discuss solutions that might benefit, among other things, the MCR process, without discussing the implications of the approach on the code review process itself (e.g., defect prediction). As a result, we excluded such papers and added exclusion criterion E2. We conducted a second pilot study on 20 additional papers using the revised criteria. As a result, we achieved better understanding of the selection criteria. We decided therefore to distribute the selection of the remaining papers among all three authors equally. In cases where more that one version of a paper was available (e.g. a conference paper and a journal extension), we selected the most recent version.\\
\textbf{SMS2022 Selection:} After conducting the survey, we extended our mapping study. We conducted another pilot study on 30 random studies to evaluate if our selection (inclusion/exclusion) criteria need updating. We were consistent to a large degree in the pilot study and found we were able to make the selection based on our initial selection criteria. Out of 30 papers, we disagreed on six papers. By disagreement, we mean that one of the authors decided to exclude the paper while the other author included the paper. We agreed to revisit the papers before making the final decision. After revisiting the papers, we eventually decided to exclude them. We divided the remaining papers and continued the selection process independently. 

\paragraph{Keywording}
\textbf{SMS2018 Keywording: }The goal of keywording the abstracts is to extract and classify the main contribution of the selected primary studies. We performed the keywording during the selection process. The keywording process resulted in a preliminary grouping into six main, and 15 studied sub-aspects in MCR research, published in our previous work~\cite{badampudi_modern_2019}. The result of the keywording process was used to create initial themes and to identify extraction items. \\
\textbf{SMS2022 Keywording:} We used the identified themes in the SMS2018 study to classify the primary studies in SMS2022; therefore, we did not need any keywording process. 

\paragraph{Data extraction}
\begin{table}
\centering
\scriptsize
\caption{Data extraction form}
\label{tab:dfe}
\begin{tabular}{p{0.07\textwidth}p{0.15\textwidth}p{0.70\textwidth}} 
\toprule
 \textbf{Type} & \textbf{Name} & \textbf{Description}\\
\midrule
\multirow{3}{*}{Meta-data} & Authors & The authors of the paper.\\
        & Publication type  & Conference or journal.\\
        & Year & Publication year of the paper.\\
        & Citations & To evaluate the research impact we count the paper's Google Scholar citations (as of June 2021).\\
\midrule
\multirow{4}{*}{Content} & Research facet & Based on Wieringa's~\cite{Wieringa:2005:REP:1107677.1107683} classification: evaluation research, solution proposals, validation research, philosophical papers and experience papers.\\
        & Rigor and relevance &  To evaluate the rigor and relevance~\cite{ivarsson2011method} of the paper we extract the following aspects. (a) Rigor: context, study design and validity threats. (b) Relevance: subjects involved in the study, context, scale and research method.\\
        & Main contribution & A summary (either verbatim from the paper or formulated by the extractor) of the paper's main contribution and results.\\
\bottomrule
\end{tabular}
\vspace{-4mm}
\end{table}

We extracted the data items shown in Table~\ref{tab:dfe} in both mapping studies.\\
\textbf{SMS2018 Data extraction:} We used the preliminary results (studied aspects in MCR research) as an input to extract additional items related to the main contribution of the primary studies. For example, we extracted the "techniques used" and "purpose" in papers providing tool support and the "techniques used" and "selection criteria" as additional extraction items for papers providing reviewer recommendation support. 
Similar to the paper selection, we planned to distribute the data extraction work among the three authors of this paper. Hence, to align our common understanding of the data extraction form, we conducted two pilots. \emph{Data extraction pilot 1:} Before starting the pilot process, we reviewed the rigor and relevance criteria provided by Ivarsson and Gorschek~\cite{ivarsson2011method}. Assessing the research method in the relevance dimension of primary studies that analyse repositories was not straightforward. For evaluating the relevance of the research methods, we agreed that the tools/solutions or findings from primary studies should be validated or evaluated by the users in the field to get a high score. Results based on solely analyzing repositories are not enough; other sources such as interviews/surveys should be conducted. Once we had an understanding of the criteria we piloted the extraction process using six primary studies where each author independently extracted the data. All the extracted items were consistent among the authors, except for the type of the subjects in the relevance, and study design in the rigor dimension. For study design, the extracted values were different for one paper and the difference was resolved in a meeting. For the type of subjects, we decided to give 0 to subjects if no subjects are involved or if the main findings of the paper are not discussed with any subjects. It is important to evaluate the relevance of the findings with users involved in code review; therefore, we give a high score when subjects are involved to corroborate the findings. We calculated the inter rater agreement (IRA) for each of the rigor and relevance aspects. For pilot 1 the average IRA for all aspects was 82\%. \emph{Data extraction pilot 2:} Based on the updated description of the rigor and relevance criteria, we conducted another pilot study on 11 papers. In the second pilot study, there was higher agreement in the extracted items. In particular, the inter rater agreement for relevance of subjects was increased considerably. For pilot 2 the average IRA for all aspects increased to 88\%.

After piloting the data extraction, we divided the data extraction of included studies until year 2018 with 20\% overlap among the authors. The average IRA between the first and second authors was 95\%, and 86\% between first and third author. For the second and third author the IRA was 75\% however, this low percentage is mainly due to a conflict in one paper. \\
\textbf{SMS2022 Data extraction:} We used the same extraction form as in SMS2018. We independently conducted the extraction of the studies identified in SMS2022 as we had a good understanding of the extraction items.


\paragraph{Data analysis}
\textbf{SMS2018 Data analysis:} Using a deductive approach, we used thematic analysis~\cite{6092576} to categorize the primary studies into themes. The main contribution extracted from the primary studies was used to generate themes. For example, if the paper's main contribution is to provide solution support for reviewer selection then we assign the paper to the "solution support" theme. We divided the primary studies based on the studies aspects identified in the keywording process among all three authors to generate themes. For example, the first author identified themes for all the primary studies related to the MCR process and studies investigating source code and review comments. The second author identified themes within the primary studies providing tool support, and finally, the third author analyzed the primary studies providing reviewer recommendations. We then followed a review process where two other authors reviewed each paper classified by one author. Based on the discussions in the review process, we moved the primary studies into different themes when needed. This process continued until all three authors reached a consensus.\\
\textbf{SMS2022 Data analysis:} We mapped the contributions of the primary studies identified in SMS2022 study to existing themes. All authors were involved in the mapping process. We reviewed the mapping process to ensure that the mapping was done fairly and not forced into existing themes. 

\paragraph{Evidence briefings} Evidence briefings are a technology transfer medium that summaries the research results in a practitioner-friendly format \cite{cartaxo2016evidence}. We created evidence briefings based on the main findings of the primary studies identified in SMS2018. We provided the link to the evidence briefings at the end of the survey allowing the practitioners to read more on the themes that they find most interesting. The evidence briefings are available online~\cite{deepika_badampudi_2021_briefings}\footnote{\url{https://rethought.se/research/modern-code-reviews/} is the original link we shared with the participants of the survey. We created the Zenodo record for archival purposes.}. 

\subsection{Q-Methodology survey} \label{Qmethod}
We chose Q-Methodology as a data collection and analysis approach since we were interested in understanding the viewpoint of practitioners on the numerous modern code review research topics we have identified in our mapping study. Q-Methodology allows to collect viewpoints on concepts that might share underlying common factors, and that are brought to the surface by revealing relations between concepts instead of rating these concepts in isolation~\cite{brouwer1999q}. The factors are identified by analyzing the subjective opinions of individuals, not facts, revealing common viewpoints within the surveyed community~\cite{brown1993primer}. A strength of the Q-Methodology is that it can provide insights on the matter of study with a relatively low number of respondents, compared to conventional surveys~\cite{davis2011q}.

Q-Methodology consists of the following seven steps~\cite{zabala2016bootstrapping}: 

\emph{(1) Defining the concourse:} The concourse consists of all the statements on a particular topic, which is in our case, modern code reviews. In our systematic mapping study, we have extracted the primary study's main contributions and research objective to represent the concourse.
    
\emph{(2) Defining the Q-Set:} The Q-Set is a sample of statements drawn from the concourse. These statements are formulated in a way such that it is possible to form an opinion about the content of the statements. The statements should not be factual, i.e. should not present a truth such as "the earth revolves around the sun", but rather evoke agreement or disagreement. Using our mapping study (SMS2018), we formulated statements from the papers' research objective in the form of "It is important to investigate ... [research statement]". We consolidated overlapping research objectives into single statements, i.e. the number of statements (46) < number of primary studies (149).  All authors were responsible for creating the statements from the primary studies in SMS2018. In a review process we revised the mapping whenever necessary. We followed the same mapping process for the primary studies identified in SMS2022. 
We list the statements and the frequency we encountered the corresponding research objective in our mapping study in Table~\ref{tab:statementsfreq}.

\emph{(3) Defining the P-Set:} The P-Set represents the survey participants. We sent the survey to practitioners in our professional network who are involved in software development and the code review process. We sent the survey link to our contact persons in 17 partner companies, who distributed the survey within the company. In addition, we published the survey link on our LinkedIn profile. The participants (P-Set) rank the statements (Q-Set) on a scale of +3 to -3 representing their agreement level. The details on the P-Set, i.e. the demographic information of participants is provided in Table~\ref{Tab:Demo}.
    
\begin{figure}
        \centering
        \includegraphics[width=\textwidth/2]{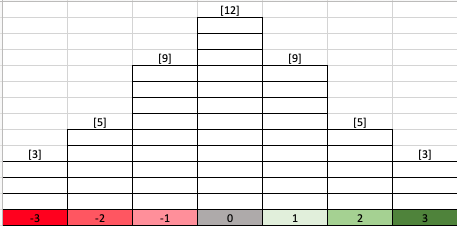}
        \caption {Q-Sort structure: The number in the square brackets represents the
number of statements per rating.}
        \label{fig:Q-SortStructure}
        \vspace{-4mm}
    \end{figure}

\emph{(4) Defining the Q-Sort structure:} The participants were asked to place a certain amount of statements in each rating. The number of statements that can be placed in each rating is shown in Figure~\ref{fig:Q-SortStructure}. For example, the participants can file 9 statements into the rating +1 and -1 rating respectively.  
    
\emph{(5) Conducting the Q-Sorting process:} In this step, the participants in the P-Set rate the statements in Q-Set. We adapted an existing tool~\cite{cartaxo2019esem} that implements the Q-Sort in a web application. After we piloted the tool with three industry practitioners, we improved the tools user interface and description of steps. The updated tool\footnote{\url{https://github.com/DeepikaBadampudi/qmethod}} and the survey\footnote{\url{https://mcrq.rethought.se/\#/step1}}, consisting of four steps, can be accessed online. In step 1, we provide an introduction to the study purpose so that the participants are aware of the importance of their input. In step 2, we ask the participants to place the statements in agree, disagree and neutral piles. In step 3, the participants place the statements in the Q-Sort structure depicted in Figure~\ref{fig:Q-SortStructure}. In this way, the participants read the statements two times; once when they put them in the agree/disagree/neutral piles and then when they file them into the Q-Sort structure. In step 4, we ask the participants to provide explanations for the placement of the six statements that they filed in the extreme agreement (+3) and disagreement (-3) piles. 
    
\emph{(6) Conducting Factor Analysis:} In the analysis, all similar viewpoints are grouped into a factor. Q-Methodology automatically classifies the viewpoints and provides a list of statements along with z-scores and factors scores for each factor. In other words, a list of statements that differentiate a viewpoint is generated along with the scores which indicate how the viewpoints differ. We elaborate on how the factors are generated in Section~\ref{Sec:differentViews}. 
    
\emph{(7) Conducting Factor Interpretation:} This last step refers to the interpretation of the factors by considering the statements and their scores in that factor and the participants' demographic information. All the statements in each factor along with the ratings were reviewed to understand the nuances of each viewpoint (factor). The first author formulated interpretations of each factor, considering the participants explanations for the statements rated with high agreement and disagreement. The interpretation was then reviewed by the second and third author. The review process resulted in minor reformulations. The factor interpretation is provided in Section \ref{Sec:differentViews}.

\section{Mapping study results}
\label{Sec:msresults}
In this section, we report on the results that answer RQ1 --- \emph{Which MCR themes have been investigated by the research community?} --- and its sub questions, based on 244 identified primary studies. 

\subsection{Research venue, type, quality, and context} \label{Sec:Quality}

\begin{figure}[b]
\centering
\begin{subfigure}[h]{.45\textwidth}
  \centering
  \includegraphics[width=.9\linewidth]{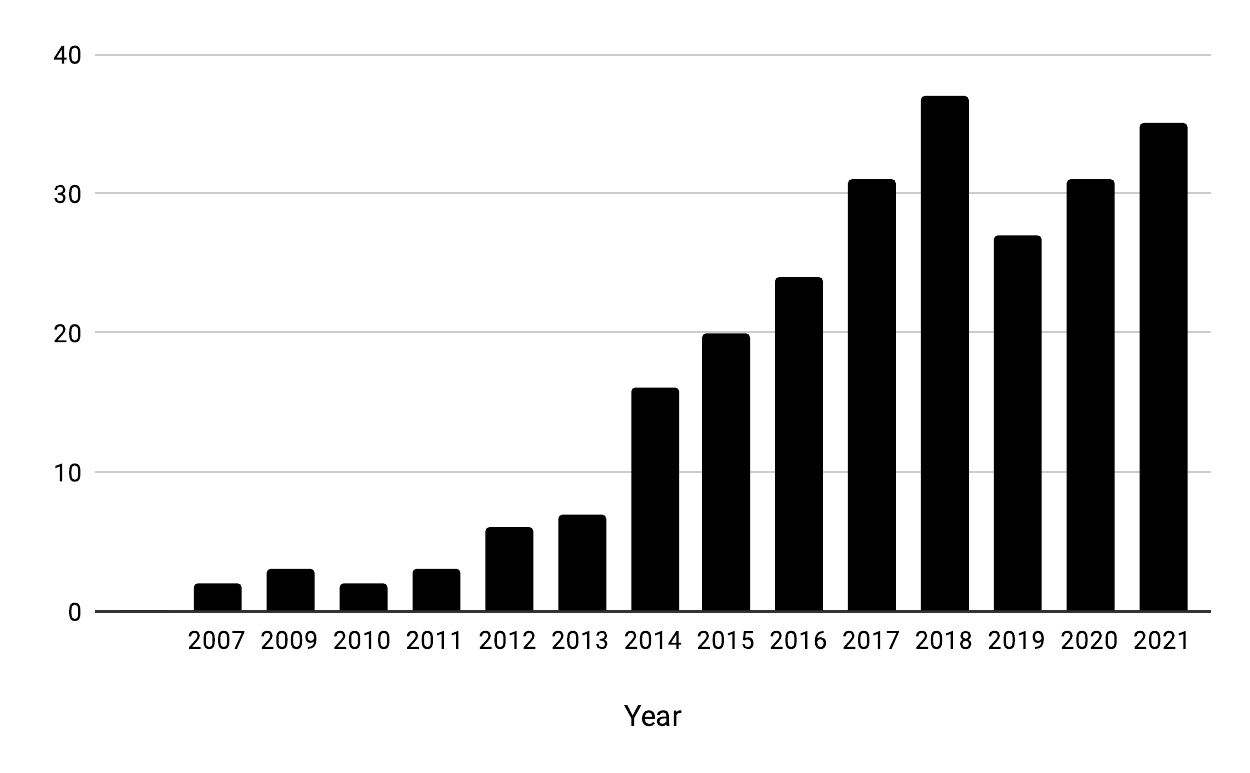}
    \caption{Year-wise count of publication types}
    \label{fig:research_year}
\end{subfigure}
\hfill
\begin{subfigure}[h]{.50\textwidth}
  \centering
  \includegraphics[width=.9\linewidth]{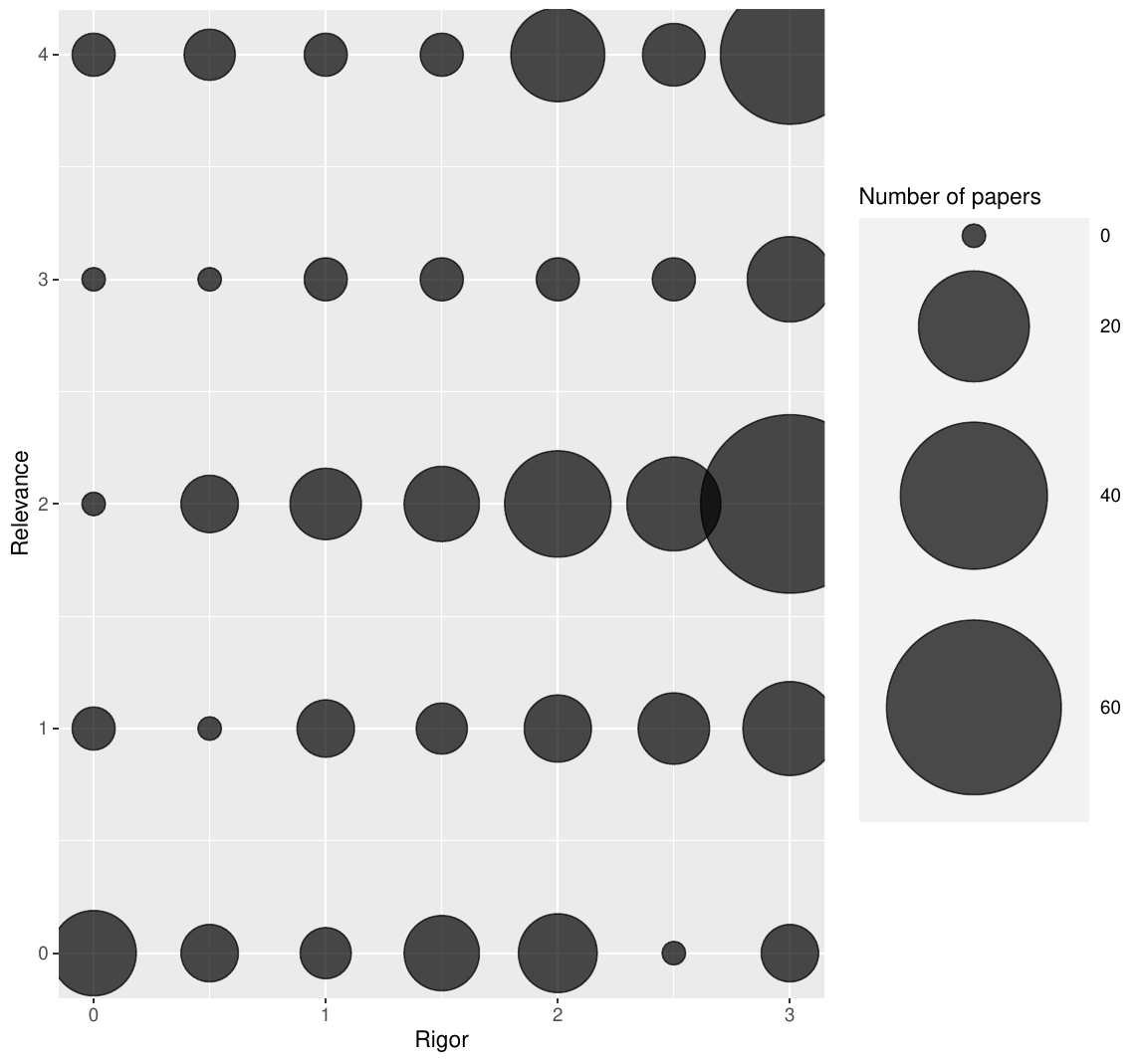}
    \caption{Total scores of rigor and relevance}
    \label{fig:RnR}
\end{subfigure}%
\caption{Frequencies of project types and research facets in primary studies}\label{fig:RnR_Year}
\end{figure}
We see a steady growth in the number of publications on modern code review as seen in Figure~\ref{fig:research_year}. In this section we provide details on the quality and the context of the primary studies. 

\textbf{Quality: Rigor and relevance -} The description of the context, study design and validity threats are used to determine the rigor. We assign a score of 1 for each rigor aspect when all details are reported, 0.5 when partial details are reported and 0 when no details are reported. The relevance is scored between 0 and 1 for the relevance of subjects involved, context, research method and scale. Once we scored each quality aspect, we calculated the total of rigor and relevance for each primary study. The maximum rigor score is 3 and relevance score 4. The total scores of rigor and relevance shown in Figure~\ref{fig:RnR}. The scores of rigor and relevance are shown on the x- and y-axes, respectively. The bubble represents the number of papers for a given score. 

As seen in Figure~\ref{fig:RnR}, the biggest bubble representing (26\%) of the primary studies has high rigor total (score = 3) and low (score = 2) relevance. Only 14\% of the primary studies have both high rigor (3) and relevance (4). Overall relevance is low because most of the primary studies did not include any human subjects in the research design. Only 32\% of the primary studies involved human subjects in their study and evaluated their findings in a realistic environment. Most primary studies analyzed repositories without triangulating the findings from other sources such as interviews or surveys. 

\begin{figure}
\centering
\begin{subfigure}[h]{.45\textwidth}
  \centering
  \includegraphics[width=.9\linewidth]{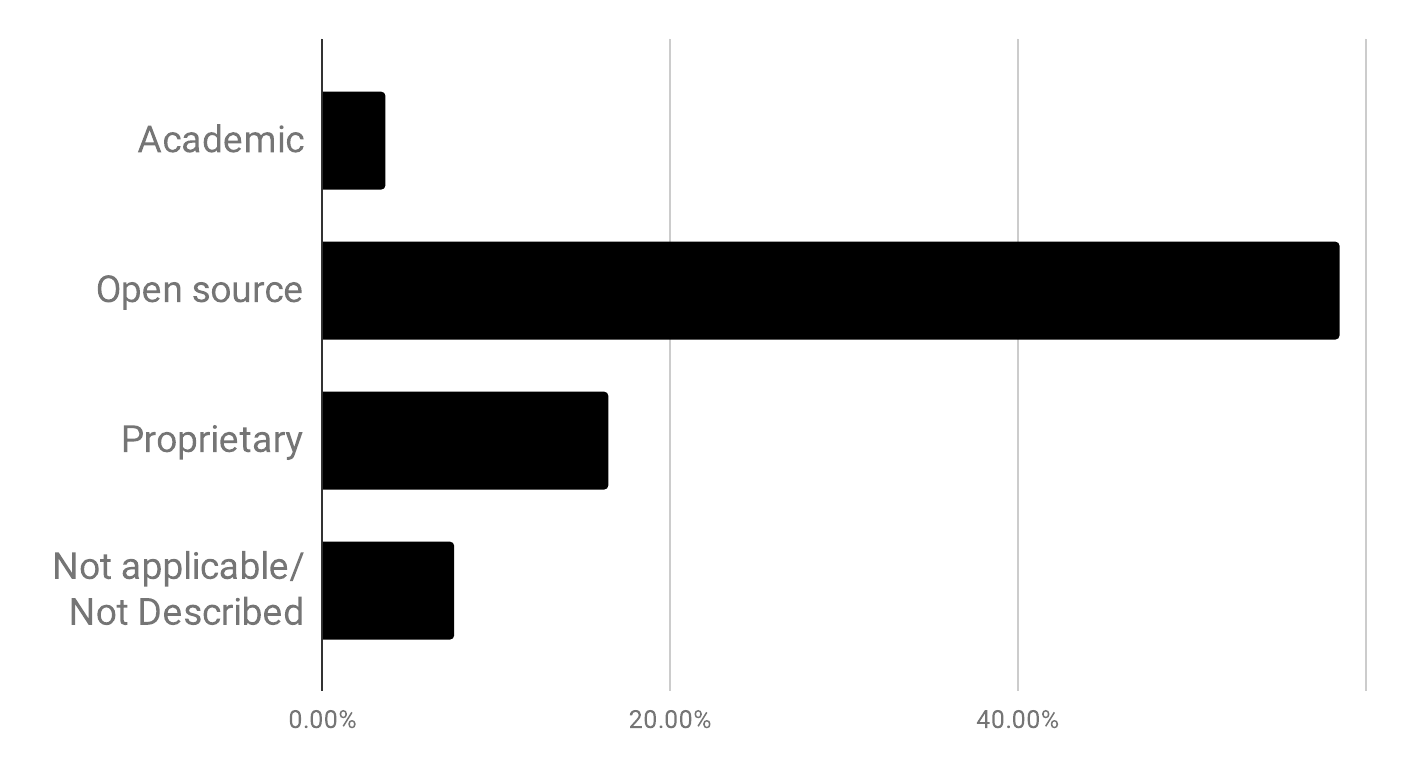}
    \caption{Projects types}
    \label{fig:ProjectTypes}
\end{subfigure}%
\hfill
\begin{subfigure}[h]{.45\textwidth}
  \centering
  \includegraphics[width=.9\linewidth]{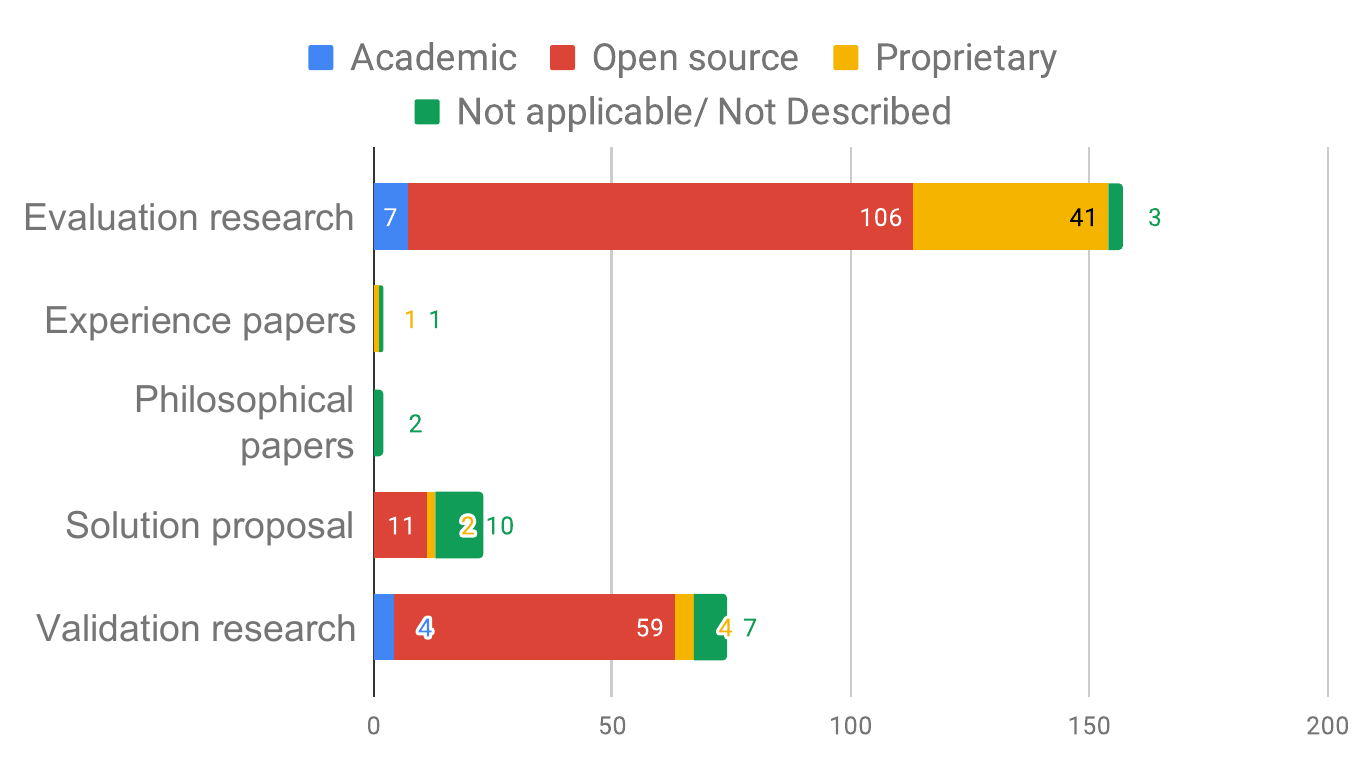}
    \caption{Research facet and project types}
    \label{fig:research_type}
\end{subfigure}
\caption{Frequencies of project types and research facets in primary studies}\label{fig:ptrf}
\end{figure}

\textbf{Context -} The type of projects investigated in the primary studies is shown in Figure~\ref{fig:ProjectTypes}. Most of the primary studies (59\%) investigated open source (OSS) projects. Less than 20\% of the primary studies investigated proprietary projects. About 10\% of the papers did not base their findings on any projects; these papers are mainly solutions papers.

Evaluation research investigates a problem or implements a solution in practice. As seen in Figure~\ref{fig:research_type}, 67\% of evaluation is done in open source projects and 26\% in proprietary projects. A small percentage (4\%) of evaluation is done in academic projects. Four papers did not describe the context of the evaluation. The same trend can be seen for validation research coverage. "Validation research investigates the properties of a solution proposal that has not been implemented in practice"~\cite{Wieringa:2005:REP:1107677.1107683}. Most of the validation is done in open source projects and only four studies are validated in a proprietary project. Solution proposals are primary studies that propose a solution and argue for its relevance. Unlike validation research, there is no thorough evaluation or validation of the solution in solution papers. One of the experience papers is based on a proprietary project, however the project type is not discussed in the other two philosophical and experience papers.

\subsection{MCR themes and contributions}\label{sec:mcrcontributions}
We grouped the 244 primary studies in five themes. In the remainder of this section, we summarize their main contributions.  

\subsubsection{SS - Support systems for code reviews} \label{Sec:Solutions} This theme includes primary studies that contribute to solutions to support the MCR process, such as review prioritization, review automation, and reviewer recommendation.\\
\textbf{Reviewer recommendations.}
A majority of papers on this theme focus on proposing tools to recommend reviewers and validate their approaches using historical data extracted from open source projects.

Most approaches recommend code reviewers based on the similarity between files modified or reviewed by each developer and the files of a new pull request (path similarity)~\citeS{Chueshev:2020,strand2020using, Siow:2020,Ye:2019,fejzer2018profile, jiang2017should, kovalenko2018does, thongtanunam2014improving, xia2015should, jiang2015coredevrec, ying2016earec,  hannebauer2016automatically, thongtanunam2015should, yang2018revrec, ouni2016search, peng2018exploring, rahman2016correct, zanjani2016automatically}. Some studies include other predictors such as previous interactions between submitter and potential reviewers~\citeS{LIAO:2020,Ye:2019, jiang2017should, jiang2015coredevrec, ying2016earec, ouni2016search}, pull request content similarity~\citeS{Nafiz:2020, jiang2017should, xia2015should, ying2016earec}, contribution to similar files~\citeS{kovalenko2018does, hannebauer2016automatically, Sulun:2019,Ye:2021a}, review linkage graphs \citeS{Hirao:2019a}, and developer activeness in a project~\citeS{Nafiz:2020, jiang2017should, kovalenko2018does, jiang2015coredevrec, yang2018revrec}. Another popular predictor to recommend code reviewers is the similarity between the content of previous and new pull requests~\citeS{jiang2017should, yang2017empirical, yu2014reviewer, xia2015should, liao2017topic, kim2018understanding, ying2016earec,xia2017hybrid}.

In one study, the authors used participants' preferences in review assignment~\citeS{wang2016code}, while in the other study, the authors combined the metadata of pull requests with the metadata associated with potential reviewers~\citeS{de2015developers}. Another study focuses on detecting and removing systematic labeling bias to improve prediction~\citeS{Tecimer:2021}. Another interesting direction is to focus recommend reviewers that will ensure code base knowledge distribution~\citeS{Mirsaeedi:2020,Rebai:2020, Chouchen:2021}. Finally, some studies have included balancing review workload as an objective~\citeS{Chouchen:2021, Zubaidi:2020, strand2020using,Asthana:2019}
 
 In relation to how the predictors are used to recommend code reviewers, many employ traditional approaches (e.g., cosine similarity), while some use machine learning techniques, such as Random Forest~\citeS{de2015developers}, Naive Bayes~\citeS{de2015developers, Tecimer:2021}, Support Vector Machines~\citeS{yu2014reviewer, jiang2015coredevrec}, Collaborative Filtering\citeS{Chueshev:2020, strand2020using}, Deep Neural Networks~\citeS{Siow:2020,Ye:2021a}, or model reviewer recommendation as an optimization problem ~\citeS{Chouchen:2021,Nafiz:2020,Rebai:2020, Zubaidi:2020, ouni2016search}.

The performance of the identified approaches varies a lot and is often measured using Accuracy~\citeS{de2015developers, yang2017empirical, thongtanunam2014improving, jiang2015coredevrec, hannebauer2016automatically, rahman2016correct, strand2020using}, Precision and Recall~\citeS{LIAO:2020,Siow:2020, fejzer2018profile, jiang2017should, yu2014reviewer, liao2017topic, ying2016earec, ouni2016search, rahman2016correct, zanjani2016automatically}, or Mean Reciprocal Rank~\citeS{strand2020using, Sulun:2019, Tecimer:2021, fejzer2018profile, xia2015should, jiang2015coredevrec, zanjani2016automatically, strand2020using}. Out of the identified studies, only a few~\citeS{strand2020using, Asthana:2019,kovalenko2018does, peng2018exploring} have evaluated code reviewer recommendation tools in live environments. Instead, the majority of the studies measures performance (accuracy, precision, recall, and mean reciprocal rank) by comparing the actual list of reviewers present in historical data with the list of developers recommended by their respective approaches.

One study focuses on identifying factors that should be accounted for when recommending reviewers~\citeS{SOARES201832}, such as the number of files and commits in a pull request, pull requester profile, previous interactions between contributors, previous experience with related code, and ownership of modified code are factors related to how code reviewers are selected.

Finally, only two studies evaluate whether reviewer recommendation really adds any value~\citeS{kovalenko2018does, strand2020using}, with mixed results.

\textbf{Understanding the code changes that need to be reviewed.}
Refactoring changes the code structure to improve testability, maintainability, and quality without changing its behavior. Supporting the review of such changes has been the focus of refactoring-aware tools. 
Refdistiller aims at detecting behavior-changing edits in refactorings~\citeS{alves2014refdistiller}. The tool uses two techniques: (a) a template-based checker that finds missing edits; and (b) a refactoring separator that finds extra edits that may change a program’s behavior. In a survey of 35 developers, they found that it would be useful to differentiate between refactored and behavior-changing code, making reviews more efficient and correct. 
ReviewFactor is a tool able to detect both manual and automated refactorings~\citeS{ge2014towards}. The evaluation of the tool showed that it can detect behavior-changing refactorings with high precision (92\%) and recall (94\%). RAID~\citeS{brito2021raid} aims at reducing the reviewers' cognitive effort by automatically detecting refactorings and visualizing information relevant for the refactoring to the reviewer. In a field experiment, professional developers reduced the number of lines they had to review for move and extraction refactorings. 
CRITICS is an interactive approach to review systematic code changes~\citeS{zhang2015interactive}. It allows developers to find changes similar to a specified template, detecting potential mistakes. The evaluation indicates that (a) six engineers who used the tool, would like to have it integrated in their review environment; and (b) the tool can improve reviewer productivity, compared to a regular diffing tool.
A study at Microsoft proposes a solution to automatically cluster similar code changes~\citeS{barnett2015helping}. The clusters are then submitted for review. A preliminary user study suggests that the understanding of code did indeed improve when the changes were clustered. Similarly, SIL identifies similar code changes in a pull request and flags potential inconsistent or overlooked changes~\citeS{ayinala2020tool}. In an inspection of 453 code changes in open source projects, it was found that up to 29\% of the changes are composite, i.e., address different concerns~\citeS{tao2015partitioning}. Decomposing large pull requests into cohesive change sets could therefore be valuable. A controlled experiment study found that change decomposition leads to fewer wrongly reported issues~\citeS{di2019effects}. ChgCutter~\citeS{guo2019decomposing} provides an interactive environment that allows reviewers to decompose code changes into atomic modifications that can be reviewed and tested individually. Professional developers suggested in a interview study that the approach helps to understand complex changes. CoRA automatically decomposes commits in a pull request into clusters of related changes (e.g. bug fixes, refactorings) and generates concise descriptions that allows users to better understand the changes~\citeS{wang2019cora}. 
Another idea to reduce review effort is to prioritize code that is likely to exhibit issues. One approach is to train a Convolutional Neural Network with old review comments and source code features to identify code fragments that require review~\citeS{staron2020using}. Similarly, CRUSO classifies to be reviewed code by identifying similar code snippets on StackOverflow and analyzing the corresponding comments and meta-data, leveraging crowd-knowledge~\citeS{sharma2019using, kapur2021using}. 

Other research looked into the order in which changed files should be presented to the reviewer to achieve an effective review process~\citeS{baum2017optimal}. The study proposes the following main principle for the ordering: related change parts should be grouped as closely as possible.
Another contribution to improve the understanding of changed code suggests identifying the so-called “salient” class, i.e. the class in which the main change was made and which affects changes in other dependent classes~\citeS{huang2018salient}. The authors hypothesize that reviews could be more efficient if the salient class would be known, making the logic of the commit easier to understand. A preliminary evaluation (questionnaire-based) with 14 participants showed that the knowledge about the salient class improves the understanding of a commit. A follow-up study with a broader evaluation confirms these results~\citeS{huang2020code}.
A similar idea is implemented in BLIMP tracer, which inspects the impact of changes on a file level, rather than on a class level~\citeS{wen2018blimp}. The tool was evaluated with 45 developers and it improved speed and accuracy of identifying the artifacts that are impacted by a code change. SEMCIA was developed to reduce noise in change impact analysis and uses semantic rather than syntactic relationships. This approach reduces false positives by up to 53\% and reduces the change impact sets considerably~\citeS{hanam2019aiding}.
MultiViewer is a code change review assistance tool that calculates metrics to better understand the change effort, risk, and impact of a change request~\citeS{wang2017multi}. 
A step further goes the approach implemented in the tool GETTY, which aims at providing meaningful change summaries by identifying change invariants through analyzing code differences and test run results~\citeS{menarini2017semantics}. With GETTY, reviewers can determine if a set of code changes has produced the desired effect. The approach was evaluated with the participation of 18 practitioners. The main finding was that GETTY substantially modified the review process to a hypothesis-driven process that led to better review comments.

Another direction of research for improving code understanding for reviews uses visualization of information. For example, ViDI supports visual design inspection and code quality assessment~\citeS{tymchuk2015code}. The tool uses static code analysis reports to identify critical areas in code, display the evolution of the amount of issues found in a review session, and allow the reviewer to inspect the impact of code changes. 
Git Thermite focuses on structural changes made to source code~\citeS{salgado2017pharo}. The tool analyzes and visualizes metadata gathered from GitHub, code metrics of the modified files, and static source code analysis of the changes in pull requests. DERT aims at complementing line-based code diff tools with a visual representation of structural changes, similar to UML but in a dynamic manner, allowing the reviewer to see an overview as well as details of the code change~\citeS{balci2021augmenting}. Similarly, STRIFFS visualizes code changes in an UML class diagram, providing the reviewer an overview~\citeS{fadhel2021striffs}. CHANGEVIZ allows developers to inspect method calls/declarations related to the reviewed code without switching context, helping to understand a change and its impact~\citeS{gasparini2021changeviz}. 
OPERIAS focuses on the problem of understanding how particular changes in code relate to changes to test cases~\citeS{oosterwaal2016visualizing}. The tool visualizes source code differences and a change’s coverage impact.
Finally, a tool was developed to improve the review process of visual programming languages (such as Petri nets)~\citeS{ragusa2018code}. It supports the code review of visual programming languages, similar to what is already possible with textual programming languages.

Meyers et al. \citeS{Meyers:2018a} developed a dataset and proposed a Natural Language-based approach to identify potential problems and elicit an active response from the colleague responsible for modifying the target code. They trained a classifier to identify acted-upon comments with good prediction performance (AUC = 0.85).

\textbf{Monitoring review performance and quality.} 
Gonz{\'a}lez-Barahona et al.~\citeS{10.1007/978-3-642-55128-4_1} have proposed to quantitatively study the MCR process, based on traces left in software repositories. Without having access to any code review tool, they analyzed changelog files, commit records, and attachments and flags in Bugzilla records to monitor the size of the review process, the involved effort, and process delay. 
A similar study focused on review messages wherein changes are first reviewed through communication in a mailing list~\citeS{izquierdo2016characterization}. They developed a series of metrics to characterize code review activity, effort, and delays~\citeS{izquierdo2017using}, which are also provided through a dashboard that shows the evolution of the review process over time~\citeS{izquierdo2018software}.
Another study looked at code reviews managed with Gerrit and proposed metrics to measure velocity and quality of reviews~\citeS{LehtonenP15}. 
Similar metrics, such as code churn, modified source code files and program complexity, were used to analyze reviewer effort and contribution in the Android open source project~\citeS{mishra2014mining}. 
Other tools to analyze Gerrit review data are ReDa, that provides reviewer, activity and contributor statistics~\citeS{thongtanunam2014reda}, and Bicho, that models code reviews as information from an issue tracking system, allowing to query review statistics with standard SQL queries~\citeS{gonzalez2014analyzing}. Finally, Codeflow Analytics aggregates and synthesizes code review metrics (over 200)~\citeS{bird2015lessons}.   

\textbf{Determining the usefulness of code reviews.} 
A study of three projects developed a taxonomy of review comments~\citeS{li2017automatic, li2017they}. After training a classifier and categorizing 147K comments, they found that inexperienced contributors tend to produce code that passes tests while still containing issues, and external contributors break project conventions in their early contributions. 
In another study, Rahman et al.~\citeS{rahman2017predicting} analyzed the usefulness of 1,116 review comments (a manual process that has also been attempted to be automatized~\citeS{pangsakulyanont2014assessing}) in a commercial system. They marked a comment as useful if it triggered a code change within its vicinity (up to 10 lines) and analyzed features of the review comment pertaining to its content and author. The results indicate that useful comments share more vocabulary with the changed code, contain relevant code elements, and are written by more experienced reviewers. Similarly, another study found that experienced reviewers are capable of pointing out broader issues than inexperienced ones~\citeS{hirao2019understanding}. The study concluded that reviewer experience and patch characteristics such as commits with large and widespread modifications drive the number of comments and words in a comment~\citeS{hirao2019understanding}. A study investigated the use of existing comments in code reviews~\citeS{spadini2020primers}. The study concluded that when the existing code review comment is about a type of bug, participants are more likely to find another occurrence of this type of bug. However, existing comments can also lead to availability bias~\citeS{spadini2020primers}. 

A study of 2,817 review comments found that only about 14\% of comments are related to software design, of which 73\% provided suggestions to address the concerns, indicating that they were useful~\citeS{zanaty2018empirical}. 
Another study investigated the characteristics of useful code reviews by interviewing seven developers~\citeS{bosu2015characteristics}. The study found that the most useful comments identify functional issues, scenarios where the reviewed code fails, and suggest API usage, design patterns or coding conventions. “Useless” comments ask how an implementation works, praise code, or point to work needed in the future. Armed with this knowledge, the researchers trained a classifier that achieved 86\% precision and 93\% recall in identifying useful comments. Applying the classifier on 1.5M review comments, they found that: (a) reviewer experience with the code base correlates with usefulness of comments, suggesting that reviewer selection is crucial, (b) the smaller the changeset, the more useful the comments, and (c) a comment usefulness density metric can be used to pinpoint areas where code reviews are ineffective (e.g. configuration and build files). 
Criticism of the above pure statistical, syntactic approaches arose as the actual meaning of comments is not analyzed~\citeS{efstathiou2018code}.

\textbf{Managing code reviews.}
Before code hosting platforms, such as Github, became popular, researchers investigated how to provide support for reviews in IDEs. SeeCode integrates with Eclipse and provides a distributed review environment with review meetings and comments~\citeS{shochat2008seecode}.
Similarly, ReviewClipse supports a continuous post-commit review process~\citeS{bernhart2010adopting}.
Scrub combines regular peer reviews with reports from static source code analyzers in a standalone application~\citeS{holzmann2010scrub}. 
Java Sniper is a web-based, collaborative code reviewing tool~\citeS{zhang2011design}. 
All these early tools have been outlived by modern code hosting and reviewing infrastructure services such as GitHub, GitLab, BitBucket, Review Board, and Gerrit. However, while these platforms provide basic code reviewing functionalities, research has also looked at improving the reviewing process in different ways~\citeS{baum2016need}. 
For example, D{\"u}rschmid~\citeS{durschmid2017continuous} suggested continuous code reviews that allow anyone to comment code they are reading or reusing, e.g., from libraries. Developers can then push questions and comments to upstream authors from within their IDE, without context switching. 
Fistbump is a collaborative review platform built on top of GitHub, providing an iteration-oriented review process that makes it easier to follow rationale and code changes during the review~\citeS{kalyan2016collaborative}. 

Fairbanks~\citeS{Fairbanks:2019} has proposed the use of DBC (Design by Contract) and highlighted how it can improve how software development teams do code reviews. When both the code author and code reviewer agree to a goal of writing code with clear contracts, they can look out for the DBC practices being followed (or not) in the code being reviewed. The author lists a few DBC examples that can be used by software development teams.

Balachandran and Vipin~\citeS{Balachandran:2020} proposed changes in the developer code review workflow to leverage online clone detection to identify duplicate code during code review. They evaluated their approach through a developer survey and learned that the proposed workflow change will increase the usage of clone detection tools and can reduce accidental clones.

Hasan et al.~\citeS{Hasan:2021} developed and evaluated an approach to measure the effectiveness of developers when doing code reviews. They defined a set of metrics and developed a model to measure code review usefulness. Their approach improved the state-of-the-art by \textasciitilde25\%. They conducted a survey with participants from Samsung and learned that the respondents found their approach useful.  

\textbf{Optimizing the order of reviews.}
Code reviewers often need to prioritize which changes they should focus on reviewing first. Many studies propose to base the review decision on the likelihood that a particular change will eventually be accepted/merged~\citeS{fan2018early, Azeem:2020a, Saini:2021}. Fan et al.~\citeS{fan2018early} proposed an approach based on Random Forest. They evaluated their approach using data from three open source projects and learned that their approach is better than a random guess. In addition to the acceptance probability, Azeem et al.~\citeS{Azeem:2020a} also considered the probability that a code integrator will review/respond to a code review request. They rank the code review requests based on both the acceptance and response probabilities, which are calculated using machine learning models. They evaluated their approach using data from open source projects, and obtained solid results. Saini and Britto~\citeS{Saini:2021} developed a Bayesian Network to predict acceptance probability. The acceptance probability is combined with other aspects, such as task type and the presence of merge conflicts, to order the list of code review requests associated with a developer. They evaluated their approach both using historical data and user feedback (both from Ericsson). They learned that their approach has good prediction performance and was deemed as useful by the users.

PRioritizer is a pull request prioritization approach that, similar to a priority inbox, sorts requests that require immediate attention to the top~\citeS{van2015automatically}. Users of the system reported that they liked the prioritization, miss however insights on the rationale for the particular pull request ordering.

Other studies looked especially at the historic proneness of files to defects to direct review efforts. A study suggests combining bug-proneness, estimated review cost, and the state of a file (newly developed or changed) to prioritize files to review~\citeS{aman20130}. The evaluation, performed on two open source projects, indicates that the approach leads to more effective reviews.  A similar approach attempts to classify files as potentially defective, based on historic instances of detected faults and features on code ownership, change request and source code metrics~\citeS{madera2017case}.

Some studies focus on predicting the time to review~\citeS{Zhang:2020a, Zhao2019ImprovingTP}. Zhao et al.~\citeS{Zhao2019ImprovingTP} focused on developing an approach that focus on time to review to prioritize review requests. In their approach, they employed a learning-to-rank approach to recommend review requests that can be reviewed quickly. They evaluated their approach through a survey with GitHub code reviewers. The survey participants acknowledge the usefulness of the approach. Zhang et al.~\citeS{Zhang:2020a} employed Hidden Markov Chains to predict the review time of code changes. They evaluated their approach using data from open source projects, with promising results.

Wang et al.~\citeS{Wang:2019a} aimed at supporting review request prioritization by identifying duplicated requests. To do so, they consider a set of features, including the time when review requests are created. They developed a machine learning model to classify whether or not a code review request is duplicated. They validated their approach using data from open source, obtaining mixed results.

\textbf{Automating code reviews.}
Gerede and Mazan~\citeS{gerede2018will} have proposed to train a classifier on whether a change request is likely to be accepted or not. Knowing in advance the likelihood of a rejected change request would reduce the review effort as those changes would not even reach the reviewing stage. They found that the change requests by inexperienced developers that involve many reviewers are the most likely to be rejected. In the same line of research, Li et al.~\citeS{Li_Heng_Yi:2019} used Deep Learning to predict a change's acceptance probability. Their approach, called DeepReview, outperformed traditional single-instance approaches.

Review Bot uses multiple static code analysis tools to check for common defect patterns and coding standard violations to create automated code reviews~\citeS{balachandran2013reducing}. An evaluation with seven developers found that they agreed to 93\% of the automatically generated comments, likely due to the lack of consistent adoption of coding standards, which were the majority of the identified defects. 
Similarly, Singh et al.~\citeS{singh2017evaluating} studied the overlap of static analyzer findings with reviewer comments in 92 pull requests from GitHub. Of 274 comments, 43 overlapped with static analyzer warnings, indicating that 16\% of the review workload could have been reduced with automated review feedback. 

A series of studies investigated the effect of bots on code reviewing practice. Wessel at al.~\citeS{Wessel:2020a} conducted a survey to investigate how software maintainers see code review bots. They identified that the survey participants would like enhancements in the feedback bots provide to developers, along with more help from bots to reduce the maintenance burden for developers and enforce code coverage. A follow-up study~\citeS{wessel2021don}, in which 21 practitioners were interviewed, identified distracting and overwhelming noise caused by review bots as a recurrent problem that affects human communication and workflow. However, a quantitative analysis~\citeS{wessel2020effects} of 1194 software projects from GitHub showed that review bots increase the number of monthly merged pull requests. It showed also that after the adoption of review bots, the time to review and reject pull requests decreased, while the time to accept pull requests was unaffected. Overall, bots seem to have a positive effect on code reviews and countermeasures to reduce noise, as discussed by Wessel et al.~\citeS{wessel2021don}, can even improve that effect.

CFar has been used in a production environment resulting in: (a) enhanced team collaboration as analysis comments were discussed; (b) improved productivity as the tool freed developers from providing feedback about shallow bugs; (c) improved code quality since the flagged issues were acted upon; and (d) the automatic review comments were found useful by the 98 participating developers~\citeS{henley2018cfar}.

Recently, researchers have invested in using Deep Learning to aiming at code review automation~\citeS{Li_Heng_Yi:2019,Shi_Li_Lo_Thung_Huo_2019, Ayinala:2020b}. Some studies have focused on identifying the difference between different code revisions~\citeS{Shi_Li_Lo_Thung_Huo_2019, Ayinala:2020b}, while Tufano et al.~\citeS{Tufano:2021} focused on providing an end-to-end solution, from identifying code changes to providing review comments. Finally, Hellendoorn et al.~\citeS{Hellendoorn:2021} evaluated if it is feasible at all to automate code reviews by developing a Deep Learning-based approach to identify the location of comments. They concluded that just this simple task is very challenging, indicating that a lot of research is still required before fully automated code review becomes a reality.

\textbf{Analyzing sentiments, attitudes and intentions in code reviews.}
Understanding review comments in greater detail could lead to systems that support reviewers in both formulating and interpreting the intentions of code reviews. A study on Android code reviews investigated the communicative goals of questions stated in reviews~\citeS{ebert2018communicative}, identifying five different intentions: suggestions, requests for information, attitudes and emotion, description of a hypothetical scenario, and rhetorical questions. A study at Microsoft showed that the type of a change intent can be used to predict the effort for a code review~\citeS{wang2019leveraging}. A study on the Chromium project found that code reviews with lower inquisitiveness, higher sentiment (positive or negative) and lower syntactic complexity were more likely to miss vulnerabilities in the code~\citeS{munaiah2017natural}. 

Several studies investigated how sentiments are expressed in code reviews~\citeS{ahmed2017senticr,hossain2020measuring,egelman2020predicting}. SentiCR flags comments as positive, neutral or negative with 83\% accuracy~\citeS{ahmed2017senticr} and was later compared to classifiers developed for the software engineering context. Surprisingly, it was outperformed by Senti4SD~\citeS{calefato2018sentiment}. The same investigation found that contributors often express their sentiment in code reviews, and that negative and controversial reviews lead to a longer review completion time~\citeS{el2019empirical}. A study at Google investigated interpersonal conflict and found in a survey that 26\% have at least once a month negative experiences with code reviews~\citeS{egelman2020predicting}. Furthermore, they found that rounds of a review, reviewing and shepherding time have a high recall but low precision in predicting negative experiences. Other research has focused on nonverbal physiological signals, such as electrodermal activity, stress levels, and eye movement to measure affect during code reviews. These signals were associated with increased typing duration and could be used in the future to convey emotional state to improve the communication in code reviews that are typically conducted without direct interaction~\citeS{vrzakova2020affect}. A study categorized incivility in open source code review discussions~\citeS{ferreira2021shut}. The results indicate that more than half (66.66\%) of the non-technical emails included uncivil features. Frustration, name calling, and impatience are the most frequent features in uncivil emails. The study also concluded that sentiment analysis tools cannot reliably capture incivility. In a study of six open source projects, men expressed more sentiments (positive/negative) than females~\citeS{paul2019expressions}. 

\textbf{Code reviews on touch enabled devices.}
Müller et al.~\citeS{muller2012approach} have proposed to use multi-touch devices for collaborative code reviews, in an attempt to make the review process more desirable. The approach provides visualizations, for example to illustrate code smells and metrics. 
Other researchers have compared reviews performed on the desktop and on mobile devices~\citeS{frkacz2017experimental}. In an experiment, they analyzed 2,500 comments, produced by computer science students and found that: (a) the reviewers on the mobile device found as many defects as the ones on the desktop; and (b) seemed to pay more attention to details.

\textbf{Other solutions.}
Some primary studies propose an initial proof of concept approaches for different purposes: to automatically classify commit messages as clean or buggy~\citeS{lal2017code}, to eliminate stagnation of reviews and to optimize the average duration of open reviews~\citeS{viviani2016removing}, use of interactive surfaces for collaborative code reviews~\citeS{raab2011collaborative}, and to link code reviews to code changes~\citeS{paixao2018crop}. In addition, a study~\citeS{lipcak2018large} compares two reviewer recommendations algorithms, concluding that recommendation models needs to be trained for a particular project to perform optimally.\\

\textbf{Links to tools and databases available reported in the primary studies: } We extracted the links to tools and databases reported in the primary studies providing solutions to support modern code reviews. Only a few primary studies provide links to the proposed tools or databases used in the studies. Most of the proposed solutions were for supporting reviewer recommendations. However, only two out of 36 solutions provided links to the tools, and seven primary studies provided links to the database they used in their studies. We observed most reporting of links (17/28) to tools and databases for the primary studies providing support to understand changes that need to be reviewed. The complete list of the links to the tools and databases, along with the purpose of the links, is available in our online repository \cite{deepika_badampudi_2022_7066821}.

\subsubsection{HOF - Human and organizational factors} 
This theme includes primary studies that investigate the reviewer and/or contributor as subject, for example, reviewer experience and social interactions. Studies that contribute to the human factors (e.g. experience) and organizational factors (e.g. social network) are categorized into this theme. \\
\textbf{Review performance and reviewers’ age and experience.}
The most investigated topic in this theme is the relation between the reviewers' age and experience on the review performance. Studies found that reviewer expertise is a good indicator of code quality~\citeS{24, Rigby:2012, Czerwonka:2015, 2, 27, Kononenko:2016}. In addition, studies found that reviewers' experience~\citeS{Kononenko:2016} and developers' experience~\citeS{Kovalenko:2018, Baysal:2016} influence the code review outcome such as review time and patch acceptance or rejection. A study investigated human factors (review workload and social interactions) that relate to the participation decision of reviewers~\citeS{ruangwan2018impact}. The results suggest that human factors play a relevant role in the reviewer participation decision. Another study investigated if age affects reviewing performance~\citeS{Murakami:2017}. The study compared students in their 20’s and 40’s showed no difference based on age or development experience. Finally, there exists some early work on harvesting reviewer experience through crowdsourcing the creation of rules and suggestions~\citeS{Ichinco:2014}.\\
{\textbf{Review performance and reviewers’ reviewing patterns and focus.}} Eye tracking has been used in several studies to investigate how developers review code. Researchers found that a particular eye movement, the scan pattern, is correlated with defect detection speed~\citeS{Uwano:2007, Sharif:2012, Begel:2018}. The more time the developer spends on scanning, the more efficient is the defect detection~\citeS{Sharif:2012}. Based on these results, researchers have also stipulated that reviewing skill and defect detection capability can be deduced from eye movement~\citeS{Chandrika:2018}. Studies compared the review patterns of different types of programmers~\citeS{hauser2020code, huang2020biases}. A study compared novice and experienced programmers and based on the eye movements and reading strategies concluded that experienced programmers grasped and processed information faster and with less effort~\citeS{hauser2020code}. When comparing the eye tracking results based on gender, a study found that men fixated more frequently, while women spent significantly more time analyzing pull request messages and author pictures~\citeS{huang2020biases}.  \\
{\textbf{Review performance and reviewers’ workload.}} The impact of workload on code reviews has been investigated from two perspectives. First, a study found that workload (measured in pending review requests) negatively impacts review quality in terms of bug detection effectiveness~\citeS{27}. Second, a study crossing several open source projects found that workload (measured in concurrent and remaining review tasks) negatively impacts the likelihood that the reviewers accepts a new review invitation~\citeS{ruangwan2018impact}.\\
{\textbf{Review performance and reviewers’ social interactions.}}
Code reviews have been studied with different theoretical lenses on social interactions. A study used social network analysis to model reviewer relationships and found that the most active reviewers are at the center of peer review networks~\citeS{Yang:2014}. Another study used the snowdrift game to model the motivations of developers participating in code reviews~\citeS{Kitagawa:2016}. They describe two motivations: (i) a reviewer has a motive of choosing a different action (review, not review) from the other reviewer, and (ii) a reviewer cooperates with other reviewers when the benefit of review is higher than the cost. A study found that past participation in reviews on a particular subsystem is a good predictor for accepting future review invitations~\citeS{ruangwan2018impact}. Similarly, another study looking at review dynamics, found the amount of feedback at patch has received is highly correlated with the likelihood that the patch is eventually voted to be accepted by the reviewer~\citeS{thongtanunam2020review}.\\
{\textbf{Review performance and reviewers’ understanding of each other's comments.}} A study on code reviews investigated if reviewers’ confusion can be detected by humans and if a classifier can be trained to detect reviewers’ confusion in review comments~\citeS{ebert2017confusion}. The study concludes that while humans are quite capable of detecting confusion, automated detection is still challenging. Ebert et al.~\citeS{ebert2021exploratory} identified causes of confusion in the code: the presence of long or complex code changes, poor organization of work, dependency between different code changes, lack of documentation, missing code change rationale, and lack of tests. The study also identified the impact of confusion and strategies to cope with confusion. \\
{\textbf{Review performance and reviewers’ perception of code and review quality.}} A survey study conducted among reviewers identified factors that determine their perceived quality of code and code reviews~\citeS{Kononenko:2016}. High quality reviews provide clear and thorough feedback, in a timely manner, by a peer with a supreme knowledge of the code base, strong personal and interpersonal qualities. Challenges to achieve high quality reviews are of technical (e.g. familiarity with the code) and personal (e.g. time management) nature.\\
{\textbf{The difference between core and irregular contributors and reviewers.}} Studies investigated the difference between core and irregular contributors and reviewers in terms of review requests, frequency and speed~\citeS{Baysal:2012b, 18, Kerzazi:2016, Bosu:2012, pinto2018gets}. A study found that contributions from core developers were rejected faster (to speed-up development), while contributions from casual developers were accepted faster (to welcome external contributions)~\citeS{Baysal:2012b}. Similar observations were made in other studies~\citeS{18, Kerzazi:2016}, while Bosu and Carver~\citeS{Bosu:2012} found that top code contributors were also the top reviewers. A study explored different characteristics of the patches submitted to company-owned OSS project and found that volunteers face 26 times more rejections than employees~\citeS{pinto2018gets}. In addition, the review of patches submitted by volunteers have to wait, on average, 11 days whereas employees wait two days on average. 
Studies also investigated the acceptance likelihood of core and irregular contributors~\citeS{18, Hellendoorn:2015}. Bosu and Carver~\citeS{18} found that core contributors are more likely to have their changes accepted to the code base than irregular contributors. A potential explanation for this observation was found in another study~\citeS{Hellendoorn:2015}, showing that rejected code is significantly different (due to different code styles) to the project code than accepted code. More experienced contributors submit code that is more compliant to the project’s code style. 
A study investigated the consequences of disagreement between reviewers who review the same patch~\citeS{23}. The study found that more experienced reviewers are more likely to have a higher level of agreement than less experienced reviewers. A study investigating the career paths of contributors (from non-reviewer, i.e. developer, to reviewer, to core reviewer) found that (a) there is little movement between the population of developers and reviewers, (b) the turnover of core reviewers is high and occurs rapidly, (c) companies are interested in having core reviewers in their full-time staff, and (d) being a core reviewer seems to be helpful in achieving a full-time employment in a project~\citeS{Wesel:2017}.\\
{\textbf{The effect of the number of involved reviewers on
code reviews.}} A study found that the more the developers are involved in the discussion of bugs and their resolution, the less likely the reviewers are to miss potential problems in the code~\citeS{27}. The same holds not true for reviewer comments: surprisingly, the studied data indicates that the more reviewers participate with comments on reviews, the more likely they miss bugs in the code they review. Another study also made a counter-intuitive observation: files vulnerable to security issues tended to be reviewed by more people~\citeS{24}. One reported explanation is that reviewers get confused about what their role in the review is if there are many reviewers involved (diffusion of responsibility). Similar results were found in a study of a commercial application: the more reviewers are active, the less efficient the review and the lower the comment density~\citeS{Santos:2017}. In a study including both open source and commercial projects, it was observed that it is general practice to involve two reviewers in a review~\citeS{Rigby:2013}.\\
{\textbf{Information needs of reviewers in code reviews.}} A study identified the following information need categories: alternative solutions and improvements, correct understanding, rationale, code context, necessity, specialized expertise, splitability of a change~\citeS{pascarella2018information}. The authors of the study find that some of the information needs can be satisfied by current tools and research results, but some aspects seem not to be solved yet and need further investigation. Studies investigated the use of links in review comments~\citeS{jiang2017empirical,wang2021understanding}. A case study of the OpenStack and Qt projects indicated that the links provided in code review discussion served as an important resource to fulfill various information needs such as providing context and elaborating patch information~\citeS{wang2021understanding}. Jiang et al.~\citeS{jiang2017empirical} found that 5.25\% of pull requests in 10 popular open source projects have links. The authors conclude that pull requests with links have more comments, more commenters and longer evaluation time. Similar results were found in a study of three open source projects~\citeS{wang2021automatic} where patches with links took longer to review. The study also finds combining two features (i.e., textual content and file location) to be effective in detecting patch linkages. Similarly machine learning classifiers can be used to automate patch linkages~\citeS{Hirao:2019a}.\\
\vspace{-0.5cm}
\subsubsection{IOF - Impact of code reviews on product quality and human aspects} This theme includes primary studies that investigate the impact of code reviewers on artefacts such as code, design and human aspects such as attitude and understanding.\\
\textbf{The impact of code reviews on defect detection or repair.} A study showed that unreviewed commits have over twice as many chances of introducing bugs than reviewed commits~\cite{5}. Similarly, observations from another study show that both defect-prone and defective files tend to be reviewed less rigorously in terms of review intensity, participation and time than non-defective files~\citeS{12}.

Another study has investigated how code review coverage (the proportion of reviewed code of the total code), review participation (length and speed of discussions) and reviewer expertise affect post-release defects in large open source projects~\citeS{14}. The findings suggest that reviewer participation is a strong indicator for defect detection ability. While high code review coverage is important, it is even more important to monitor the participation of reviewers when making release decisions and select reviewers with adequate expertise on the specific code. However, these findings could not be confirmed in a replication study~\citeS{krutauz2020code}. They found that review measures are neither necessary nor sufficient to create a good defect prediction model. The same conclusions we confirmed in a project of proprietary software~\citeS{9}. In their context, other metrics such as the proportion of in-house contributions, the measure of accumulated effort to improve code changes and the rate of author self-verification contributed significantly to defect proneness~\citeS{9}.

Defective conditional statements are often the source of software errors. A study~\citeS{7} found that negations in conditionals and implicit indexing in arrays are often replaced with function calls, suggesting that reviewers found that this change leads to more readable code.\\
\textbf{The impact of code reviews on code quality.} Studies were conducted to find the problems fixed by code reviews. A study concluded that 75\% of the defects identified during code review are evolvability type defects~\citeS{10}. They also found that code review is useful in improving the internal software quality (through refactoring). Similarly, other studies~\citeS{beller2014modern, mantyla2008types} found that 75\% of changes are related to evolvability and only 25\% of changes are related to functionality. 

Studies investigated the impact of code reviews on refactoring. A study on 10 JAVA OSS projects found the most frequent changes in MCR commits are on code structure (e.g., re-factorings and reorganizations), and software documentation~\citeS{panichella2020empirical}. An investigation of 1,780 reviewed code changes from 6 systems pertaining to two large
open-source communities found that refactoring is often mixed with other changes such as adding a new feature~\citeS{paixao2020behind}. In addition, developers had explicit intent of refactoring only in 31\% of review that employed refactoring~\citeS{paixao2020behind}. An empirical study on refactoring-inducing pull requests found that 58.3\% presented at least one refactoring edit induced by code review~\citeS{coelho2021empirical}. In addition, Beller et al.~\citeS{beller2014modern}  found that 78-90\% of the triggers for code changes are review comments. The remaining 10–22\% are “undocumented”. Another study showed that reviewed commits have significantly higher readability and lower complexity. However, no conclusive evidence was reported on coupling~\cite{5}.

A study of Openstack patches found higher code conformance of a patch after being reviewed than a patch that was first submitted~\citeS{sri2021does}.
An investigation on the impact of code review on coding convention violation found that convention violations disappear after code reviews. However, only a minority of the violations were removed because they were flagged in a review comment~\citeS{han2020does}.
The comparison of cost required to produce quality programs using code reviews and pair programming showed that code reviews costs 28\% less compared to pair programming~\citeS{swamidurai2014investigating}. \\

\textbf{The impact of code reviews on detection or fixes of security issues.} According to a study~\citeS{13} code review leads to the identiﬁcation and fixes of different vulnerability types. The experience of reviewers regarding vulnerability issues is an important factor in finding security related problems, as a study indicates~\citeS{24}.  
Another large study~\citeS{thompson2017large} also has similar findings. The results indicate that code review coverage reduces the number of security bugs in the investigated projects. 
A study looked into the language used in code reviews to find if the linguistics characters could explain developers missing a vulnerability~\citeS{munaiah2017natural}. The study found that code reviews with lower inquisitiveness (fewer questions per sentence), higher positive or negative sentiment, lower cognitive load and higher assertions are more likely to miss a vulnerability. A study investigated the security issues identified through code reviews in an open source project~\citeS{di2016security}. They found that 1\% of reviewers’ comments are security issues. Language-specific  issues (e.g., C++ issues and buffer overflows) and domain-specific ones (e.g., such as Cross-Site Scripting) are often missed security issues and initial evidence indicates that reviews conducted by more than 2 reviewers are more successful at finding security issues. Another online study on freelance developers' code review process has similar findings indicating that developers did not focus on security in their code reviews~\citeS{danilova2021code}. However, the results showed that prompting for finding security issues in code reviews significantly affects developers' identification of security issues. A study found the relevant factors in successful identification of security issues in code reviews~\citeS{paul2021security}. The results indicate that the probability of security issues identification decreases with the increase in review factors such as number of reviewer's prior reviews, and number of review comments authored on a file during the current review cycle. In addition, the probability of security issues identification increases with review time, number of mutual reviews between the code author and a reviewer, and a reviewer’s number of prior reviews of the file.\\
\textbf{The impact of code reviews on software design.} A study~\citeS{8} found that high code review coverage can help to reduce the incidence of anti-patterns such as Blob, Data class, Data clumps, Feature envy and Code Duplication in software systems. In addition, the lack of participation (length and speed of discussions) during code reviews has a negative impact on the occurrence of certain code anti-patterns. 
Similarly, a study~\citeS{32} specifically looked for the occurrences of review comments related to five code smells (Data Clumps, Duplicate Code, Feature Envy, Large Class and Long Parameter List) and found that the code review process did identify these code smells. An empirical study of code smells in code reviews in two most active OpenStack projects (Nova and Neutron) found that duplicated code, bad naming, and dead code are the most frequently identified smells in code reviews~\citeS{han2021understanding}. Another investigation of 18,400 reviews and 51,889 revisions found that 4,171 of the reviews led to architectural changes, 731 of which were significant changes~\citeS{paixao2019impact}.

The impact of code reviews on design degradation is investigated in two studies~\citeS{uchoa2020does,uchoa2021predicting}. A study on code reviews in OSS projects found that certain code review practices such as long discussions and reviewers' disagreements can lead to design degradation~\citeS{uchoa2020does}. To prevent design degradation, it is important to detect design impactful changes in code reviews. A study found that technical features (code change, commit message, and file history dimensions) are more accurate than social ones in predicting (un)impactful changes~\citeS{uchoa2021predicting}.  \\
\textbf{The impact of code reviews on teams' understanding of the code under review.} An interview study~\citeS{spohrer2013peer}, found that code reviews help to improve the team’s understanding of the code under review. In addition, code review may be a valuable addition to pair programming, particularly for newly established teams~\citeS{spohrer2013peer}. Similarly, a survey of developers and a study of email threads found that developers find code review dialogues useful for understanding design rationales~\citeS{36}. 
Another survey of developers~\citeS{bosu2016process} found code reviews to help in knowledge dissemination. This was also found in a survey of reviewers that code review promotes collective code ownership~\citeS{16}. However, Caulo et al.~\citeS{caulo2020knowledge} were not able to capture the positive impact of code review in knowledge translation among developers. The authors contribute the negative results to fallacies in their experiment design and notable threats to validity.\\
\textbf{The impact of code reviews on peer impression in terms of trust, reliability, perception of expertise, and friendship.} A survey of open source contributors~\citeS{bosu2013impact} found that there is a high level of trust, reliability, and friendship between open source software projects’ peers who have participated in code review for some time. Peer code review helped most in building a perception of expertise between code review partners~\citeS{bosu2013impact}. Similarly, another survey~\citeS{bosu2016process} found that the quality of the code submitted for review helps reviewers form impressions about their teammates, which can influence future collaborations. \\
\textbf{The impact of code reviews on developers’s attitude and motivation to contribute.} An analysis of two years of code reviews showed that review feedback has an impact on contributors becoming long-term contributors~\citeS{15}. Specific feedback such as “Incomplete fix” and “Sub-optimal solution” might encourage contributors to continue to work in open source software projects~\citeS{15}. 
Similarly, a very large study found that negative feedback has a significant impact on developers’ attitude~\citeS{25}. Developers might not contribute again after receiving negative feedback and this impact increases with the size of the project~\citeS{25}. 

\vspace{-0.5cm}
\subsubsection{CRP - Modern code review process properties}  
This theme includes primary studies investigating how and when reviews should be conducted and characteristics such as review benefits, motivations, challenges and best practices.\\
\textbf{When should code reviews be performed?}
Research shows that code reviews in large open source software projects are done in short intervals~\citeS{Rigby:2012, Rigby:2013}. In particular large and formal organizations can benefit from creating overlap between developers' work, which produces invested reviewers, and from increasing review frequency~\citeS{Rigby:2012}.

\textbf{What are the benefits of code reviews besides finding defects?}
A study on large open source software projects found that code reviews act as a group problem-solving activity. Code reviews support team discussions of defect solutions~\citeS{Rigby:2012, Rigby:2013}. The analysis of over 100,000 peer reviews found that reviews also enable developers and passive listeners to learn from the discussion~\citeS{Rigby:2012, Rigby:2013}. A similar observation was made in a survey of 106 practitioners, where, besides knowledge sharing, the development of cognitive empathy was identified as a benefit of code reviews~\citeS{cunha2021code}. 

\textbf{How are review requests distributed?}
Research found that reviews distributed via broadcast (e.g., mailing list) were twice as fast as unicast (e.g. Jira). However, reviews distributed via unicast were more effective in capturing defects~\citeS{Armstrong:2017}. In the same investigation, code reviewers reported that a unicast review allows them to comment on specific code, visualize changes and have less traffic of patches circulating among reviewers. However, new developers learn the code structure faster with frequent circulation of patches among those who subscribe to broadcast reviews.

\textbf{Efficiency and effectiveness of code reviews compared to team walkthroughs.}
Team walkthroughs are often used in safety-critical projects, come however also with additional overhead. In a study that developed an airport operational database, the MCR process was compared with a walkthrough process~\citeS{Bernhart:2012}. The authors suggest to adopt MCR to ensure coverage while adapting the formality to the criticality of the item under review. Over-the-shoulder (OTS) reviews are synchronous code reviews where the author leads the analysis. A study compared in an experiment OTS with tool-assisted (TA), asynchronous, code reviews. It was found that OTS generates higher quality comments about more important issues, and better supports knowledge transfer, while TA generates more comments~\citeS{jureczko2020code}.

\textbf{Mentioning peers in code review comments.}
A study explored the use of @-mentions, a social media tool, in pull requests~\citeS{Zhang:2014}. The main findings were that @-mentions are used more frequently in complex pull requests and lead to shorter delays in handling pull requests. Another study investigated which socio-technical attributes of developers are able to predict @-mentions. It found that a developers visibility, expertise and productivity are associated with @-mentions, while, contrary to intuition, responsiveness is not~\citeS{kavaler2019whom}. Generalizing the idea of @-mentions, other researchers investigated to what information objects to stakeholders refer to in pull request discussions. Building taxonomies of reference and expression types, they found that source code elements are most often referred to, even though the studied platform (GitHub) does not provide any support in creating such links (in contrast to references to people or issue reports)~\citeS{chopra2021alex}.    

\textbf{Test code reviews.}
Observations on code reviews found that the discussions on test code are related to testing practices, coverage, and assertions. However, test code is not discussed as much as production code~\citeS{Spadini:2018}. When reviewing test code, developers face challenges such as lack of testing context, poor navigation support (between test and production code), unrealistic time constraints imposed by management, and poor knowledge of good reviewing and testing practices by novice developers~\citeS{Spadini:2018}. Test-driven code review (TDR) is the practice of reviewing test code before production code and studied in a controlled experiment and survey~\citeS{spadini2019test}. It was found that the practice does not change review comment quality nor the overall amount of identified issues. However, more test issues were identified on the expense of maintainability issues in production code. Furthermore, in a survey it was found that reviewing tests was perceived as having low importance and lacking tool support.

\textbf{Decision-making in the code review process.}
The review process and the resulting artifacts are an important source of information for the integration decision of pull requests. In a qualitative study limited to two OSS projects, it was found that the common, most frequent reason for rejection is unnecessary functionality~\citeS{gottigundala2021qualitatively}. In a quantitative study of 4.8K GitHub repositories and 1M comments, it was found that there are proportionally more comments, participants and comment exchanges in rejected than in accepted pull requests~\citeS{golzadeh2019effect}. Another aspect of decision-making in code reviews is multi-tasking. It was observed that reviewers participating simultaneously in several pull requests (which happens in 62\% of the 1.8M studied pull requests) increase the resolution latency~\citeS{hu2019multi}.  
MCR processes often contain a voting mechanism that informs the integrator about the community consensus about a patch. The analysis of a project showed that integrators use patch votes only as a reference and decide in 40\% of the cases against the simple majority vote~\citeS{21}. Still, patches that receive more negative than positive votes are likely to be rejected.  

\textbf{Comparison of pre-commit and post-commit reviews.}
In change-based code reviews, one has the choice to perform either pre-commit or post-commit reviews. Researchers have created and validated a simulation model, finding that there are no differences between the two approaches, in most cases~\citeS{baum2017comparing}. In some cases, post-commits were better regarding cycle time and quality. For pre-commit reviews, the review efficiency was better. 

\textbf{Strategies for merging pull requests.}
A survey of developers and analysis of data from a commercial project found that pull request size, the number of people involved in the discussion of a pull request, author experience, and their affiliation are significant predictors of review time and merge decisions~\citeS{Kononenko:2018}. It was found that developers determine the quality of a pull request by the quality of its description and complexity, and the quality of the review process by the feedback quality, test quality, and the discussion among developers~\citeS{Kononenko:2018}.

\textbf{Motivations, challenges and best practices of the code review process.}
Several studies have been conducted to investigate benefits and challenges of modern code reviews. An analysis found that improving code, finding defects and sharing knowledge were the top three out of nine identified benefits associated with code reviews~\citeS{MacLeod:2018}. Similar studies identified knowledge sharing~\citeS{Sadowski:2018, cunha2021code}, history tracking, gatekeeping, and accident prevention as benefits of code reviews~\cite{Sadowski:2018}. 
Challenges such as receiving timely feedback, review size and managing time constraints were identified as the top three out of 13 identified challenges~\citeS{MacLeod:2018, Bacchelli2013}. Challenges such as geographical and organizational distance, misuse of tone and power, unclear review objectives and context were also identified~\citeS{Sadowski:2018}. In the context of refactoring, a survey found that changes are often not well documented, making it difficult for reviewers to understand the intentions and implications of refactorings~\citeS{alomar2021refactoring}.
The best practices for code authors include writing small patches, describing and motivating changes, select appropriate reviewers and being receptive towards reviewers’ feedback~\citeS{MacLeod:2018}. The code reviewers should provide timely and constructive feedback through effective communication channels~\citeS{MacLeod:2018}. Code reviews are a well established practice in open source development (FOSS). An interview study~\citeS{alami2019does} set out to understand why code review works in FOSS communities and found that 1) negative feedback is embraced as a mean for a positive opportunity for improvement and should not be reduced nor eliminated; 2) the ethic of passion and care create motivation and resilience to rejection; 3) both intrinsic (altruism and enjoyment) and extrinsic (reciprocity, reputation, employability, learning opportunities) motivation are important. Another study proposes a catalog of MCR anti-patterns that describe reviewing behaviour or process characteristics that are detrimental to the practice: confused reviewers, divergent reviewers, low review participation, shallow review, toxic review~\citeS{chouchen2021anti}. Preliminary results from studying a small sample (100) of code reviews show that 67\% contain at least one anti-pattern.
\vspace{-0.3cm}
\subsubsection{ION - Impact of software development processes, patch characteristics, and tools on modern code reviews} This theme includes primary studies investigating the impact of processes (such as continuous integration), patch characteristics (such as change size, descriptions), and tools (e.g., statics analyzers) on modern code reviews.

\textbf{The impact of static analyzers on the code review process.}
A study on six open source projects analyzed which defects are removed by code reviews and are also detected by static code analyzers~\citeS{panichella_static_2015}. In addition, a study~\citeS{ueda2018impact} found that the issues raised by coding style checker can improve patch authors’ coding style to avoid the same type of issues in subsequent patch submissions. However, the warnings from static analyzers could be irrelevant for a given project or development context. To address this issue, a study~\citeS{ueda2019mining} proposed a coding convention checker that detects project-specific patterns. While most of the produced warnings would not be flagged in a review, addressing defects regarding imports, regular expressions and type resolutions before the patch submission would indeed reduce the reviewing effort. Through an experiment~\citeS{hentschel2016can}, it was found that the use of a symbolic execution debugger to identify defects during the code review process is effective and efficient compared to a plain code-based view. Another study~\citeS{liu2021learning}, proposed a static analyzer for extracting first-order logic representations of API directives which reduces the code review time. 

\textbf{The impact of gamification elements on the code review process.} Gamification mechanisms for peer code review are proposed in a study~\citeS{unkelos2016lets}. However, an experiment with gamification elements in the code review process found that there is no impact of gamification on the identification of defects~\citeS{Khandelwal:2017}. 

\textbf{The impact of continuous integration on the code review process.}
Experiments with 26,516 automated build entries reported that successfully passed builds are more likely to improve code review participation and frequent builds are likely to improve the overall quality of the code reviews~\citeS{Rahman:2017}. Similar findings were confirmed in a study~\citeS{zampetti2019study} that found that passed builds have a higher chance of being merged than failed ones. On the impact of continuous integration (CI) on code reviews, a study~\citeS{cassee2020silent} found that on average CI saves up to one review comment per pull request.

\textbf{The impact of code change descriptions on the code review process.}
Interviews with industrial and OSS developers concluded that providing motivations for code changes along with a description of what is changed reduces the reviewer burden~\citeS{ram2018makes}. Similarly, an analysis of OSS projects found that a short patch description can lower the likelihood of attracting reviewers~\citeS{thongtanunam2018review}.

\textbf{The impact of code size changes on the code review process.}
An investigation of a large commercial project with 269 repositories found that when patch size increases, the reviewers become less engaged and provide less feedback~\citeS{14}. 
An interview study with industrial and OSS developers found that code changes that are properly sized are more reviewable~\citeS{ram2018makes}. 
The size of patches negatively affects the review response time, as observed in a study on code reviews~\citeS{Baysal:2016}, and reduces the number of review comments~\citeS{Liang:2011}, and code review effectiveness as shown in a study of an OSS project~\citeS{baum2019associating}. 
Similarly, an analysis of more than 100,000 peer reviews in open source projects recommends that changes to be reviewed should be small, independent, and complete~\citeS{Santos:2017}.

\textbf{The impact of commit history coherence on the code review process.}
An interview study on industrial and OSS project developers found that the commit messages that are self-explanatory and have meaningful messages are easier to review~\citeS{ram2018makes}. In addition, interviewees suggest that the ratio of commits in a change to the number of files changed should not be high~\citeS{ram2018makes}.

\textbf{The impact of review participation history on the code review process.}
An analysis of three OSS projects found that the likelihood of attracting reviewers is higher when past changes to the modified files are reviewed by at least two reviewers~\citeS{thongtanunam2018review}. Prior patches that had few reviewers tend to be ignored~\citeS{thongtanunam2018review}. 
Another study, looking at reviews from two OSS projects found that more active reviewers have faster response times~\citeS{Baysal:2016}.

\textbf{The impact of fairness on the code review process.}
Fairness, in general, refers to the decision and allocation of resources in a way that is fair to the individuals and the group. A study~\citeS{german2018my} in an OSS project investigated different fairness aspects and recommends, besides the common aspects of fairness such as politeness, and precise and constructive feedback, to: (a) distribute reviews fairly, and (b) establish a clear procedure for how reviews are performed. A study~\citeS{furtado2020successful} investigated how contributions from different countries are treated. The study found that developers from countries with low human development face rejection the most. From the perspective of bias, a study~\citeS{9361116} investigated the benefits of anonymous code reviews. The results indicate that while anonymity reduces bias, it is sometimes possible to identify the reviewer and there are some practical disadvantages such as not being able to discuss with the reviewer. The study recommends to have a possibility to  reveal the reviewer when required. Another qualitative study~\citeS{nadri2021insights} found that there may be perceptible race bias in the acceptance of pull requests. Similarly a study investigated the impact of gender, human and machine bias in response time and acceptance rate~\citeS{huang2020biases}. The results indicate that gender identity has significant effect on response time and all participants spend less time evaluating the pull requests of women, and are less likely to accept the pull requests of machines. 

\textbf{The impact of rebasing operations in the code review process.} An in-depth large-scale empirical investigation of the code review data of 11 software systems, 28,808 code reviews and 99,121 revisions found that rebasing operations are carried out in an average of 75.35\% of code reviews of which 34.21\% operations tend to tamper with the reviewing process~\citeS{paixao2019rebasing}.
\subsubsection{Other} In this theme we have papers that investigate code review process on a generic level. Two papers classified the code review process in open source projects \citeS{asundi2007patch}, and in proprietary projects \citeS{baum2016faceted}. A study identifies the factors influencing the code review process \citeS{baum2016factors}. 

\subsection{Mapping of papers to Statements} \label{Sec:Statements} 
As explained in Section~\ref{Qmethod}, Q-Methodology relies on a set of statements that can be assessed by survey participants. We created these statements based on the primary studies' research objective. All primary studies with similar research objectives are mapped to one statement. Note that each primary study could be mapped to different statements when their research objectives are multifaceted.  We did not want to make the survey too long. Thus, we merged the statements that are closely related. In total, we generated 47 statements representing the primary studies. We do not have statements for three papers in the "other" theme as they were not as well aligned to the main themes. Similarly we do not have statements representing four short papers with very brief descriptions of solutions proposals; we aimed to include statements with at least two or more solutions per statement. We provide the statements derived from the primary studies in Table~\ref{tab:statementsfreq}. Note that we extended our mapping study from the review period until 2018 to 2021. Since we created the statements on a high level, we could map most of the new studies to the existing statements representing primary studies until 2018. However, we identified one new statement that we could not map to any existing statement. The new statement is related to investigating the impact of concurrent code changes on modern code reviews (last statement in the ION theme in Table~\ref{tab:statementsfreq}). 

\begin{table}[]
\centering
\scriptsize
\caption{Statements representing the five main themes and frequency of primary studies.\\ \small{\textbf{Note}: All statements start with prefix text mentioned in \textit{italics} under each theme.}}
\label{tab:statementsfreq}
\begin{tabular}{@{}lrlr@{}}
\toprule
\textbf{Statement}                                                                                                                                   & \multicolumn{1}{l}{\textbf{Papers}}                                                            & \textbf{Statement}                                                                                                                                                                                      & \multicolumn{1}{l}{\textbf{Papers}} \\
 \midrule
\multicolumn{2}{l}{Support Systems (SS): \textit{It is important to investigate support for...}}                                                                                                                                                                  & \multicolumn{2}{l}{\begin{tabular}[c]{@{}l@{}}Human and Organizational Factors (HOF): \\ \textit{It is important to investigate...}\end{tabular}}                                                                                                         \\
selection of appropriate code reviewers.                                                                                                             & 36                                                                                             & reviewers’ age and experience.                                                                                                                                                                          & 11                                  \\
understanding changes that need review.                                                                                                              & 28                                                                                             & reviewers’ reviewing patterns and focus.                                                                                                                                                                & 6                                   \\
automating code  reviews.                                                                                                                            & 11                                                                                              & reviewers’ understanding of each other's comments.                                                                                                                                                      & 2                                   \\

analyzing the sentiment/attitude/intention.                                                                                                          & 11                                                                                             & reviewers information needs.                                                                                                                                                                            & 5                                   \\
monitoring review performance and quality.                                                                                                           & 10                                                                                             & reviewers’ social interactions.                                                                                                                                                                         & 4                                   \\
determining the usefulness of code reviews.                                                                                                          & 9                                                                                              & effect of number of involved reviewers.                                                                                                                                                                 & 4                                   \\
managing code reviews.                                                                                                                               & 9                                                                                              & reviewers’ workload.                                                                                                                                                                                    & 2                                   \\
optimizing code review order.                                                                                                                        & 9                                                                                             & core vs. irregular -  requests, frequency and speed.                                                                                                                                                    & 5                                   \\

code reviews on touch-enabled devices.                                                                                                               & 2                                                                                              & core vs. irregular - acceptance likelihood.                                                                                                                                                             & 2                                   \\
                                                                                                                                                     & \multicolumn{1}{l}{}                                                                           & reviewers’ perception of code and review quality.                                                                                                                                                       & 1                                   \\
                                                                                                                                                     & \multicolumn{1}{l}{}                                                                           & core vs. irregular - agreement level.                                                                                                                                                                   & 1                                   \\
                                                                                                                                                     & \multicolumn{1}{l}{}                                                                           & core vs. irregular - career paths.                                                                                                                                                                      & 1                                   \\
TOTAL                                                                                                                                                & 125                                                                                           & TOTAL                                                                                                                                                                                                   & 45                                  \\
 \midrule
\multicolumn{2}{l}{\begin{tabular}[c]{@{}l@{}}Impact of code reviews on product quality and human aspects\\ (IOF): \textit{It is important to investigate the impact of code reviews on. . .}\end{tabular}}                                                       & \multicolumn{2}{l}{\begin{tabular}[c]{@{}l@{}}Modern code review process properties (CRP): \\ \textit{It is important to investigate. . .}\end{tabular}}                                                                                                  \\
code quality.                                                                                                                                        & 10                                                                                             & motivations, challenges and best practices.                                                                                                                                                             & 6                                   \\
defect detection or repair.                                                                                                                          & 6                                                                                              & decision making process                                                                                                                                                                                 & 4                                   \\
detection or fixes of security issues.                                                                                                               & 7                                                                                              & when to review                                                                                                                                                                                          & 3                                   \\
software design.                                                                                                                                     & 6                                                                                              & review benefits.                                                                                                                                                                                        & 3                                   \\
teams' understanding of the code under review.                                                                                                       & 5                                                                                              & mentioning peers in comments.                                                                                                                                                                           & 3                                   \\
peer impression.                                                             & 2                                                                                              & comparison to team walkthroughs.                                                                                                                                                                        & 2                                   \\
developers' attitude and motivation to contribute.                                                                                                   & 2                                                                                              & it is important to investigate reviews of test code.                                                                                                                                                    & 2                                   \\
                                                                                                                                                     & \multicolumn{1}{l}{}                                                                           & how requests are distributed.                                                                                                                                                                           & 1                                   \\
                                                                                                                                                     & \multicolumn{1}{l}{}                                                                           & pre and post-commit code reviews.                                                                                                                                                                       & 1                                   \\
                                                                                                                                                     & \multicolumn{1}{l}{}                                                                           & strategies for merging pull requests.                                                                                                                                                                   & 1                                   \\
TOTAL                                                                                                                                                & 39                                                                                             & TOTAL                                                                                                                                                                                                   & 27                                  \\
 \midrule
\multicolumn{2}{l}{\begin{tabular}[c]{@{}l@{}}Impact of software development processes, patch characteristics, \\ and tools on modern code reviews (ION): \textit{It is important to} \\ \textit{investigate the impact of . . . on the code review process}\end{tabular}} &                                                                                                                                                                                                         & \multicolumn{1}{l}{}                \\
static analyzers.                                                                                                                                    & 5                                                                                              &                                                                                                                                                                                                         & \multicolumn{1}{l}{}                \\
code size changes.                                                                                                                                   & 6                                                                                              &                                                                                                                                                                                                         & \multicolumn{1}{l}{}                \\
fairness on the code reviews process.                                                                                                                & 4                                                                                              &                                                                                                                                                                                                         & \multicolumn{1}{l}{}                \\
continuous integration.                                                                                                                              & 3                                                                                              &                                                                                                                                                                                                         & \multicolumn{1}{l}{}                \\
gamification elements.                                                                                                                               & 2                                                                                              &                                                                                                                                                                                                         & \multicolumn{1}{l}{}                \\
code change descriptions.                                                                                                          & 2                                                                                                                   & \multirow{6}{*}{\begin{tabular}[c]{@{}l@{}}GRAND TOTAL (some papers were classified\\  in more than one statement, hence the grand total \\ exceeds the  number of reviewed primary studies)\end{tabular}} & \multirow{7}{*}{263}                \\
review participation history.                                                                                                                        & 2                                                                                              &                                                                                                                                                                                                         &                                     \\
commit history coherence.                                                                                                                            & 1                                                                                              &                                                                                                                                                                                                         &                                     \\
concurrent code changes.                                                                                                                            & 1                                                                                              &                                                                                                                                                                                                         &                                     \\
TOTAL                                                                                                                                                & 27                                                                                             &                                                                                                                                                                                                         &                                  
\\
\bottomrule
\end{tabular}
\end{table}

\section{Survey Results}
\label{sec:surveyresults}
In this section, we report on the results that answer RQ2 --- \emph{How do practitioners perceive the importance of the identified MCR research themes?} --- based on 46\footnote{As the survey was conducted after our initial mapping study we could not include the new statement representing one primary study from the updated mapping study.} statements that we derived from the identified five main themes on MCR research.
\subsection{Demographics}
We received 28 responses in total. We excluded three respondents as they did not have any code review experience or entered invalid responses. The remaining 25 respondents work in different roles in large multinational organizations;  56\% of the participants are working in Swedish software organizations. The company name was not a mandatory field in the survey. However, approximately 70\% of the respondents provided their company name. 
Most of the respondents are from the telecommunication domain. We also received responses from practitioners working in product-based companies and IT services and consulting companies. We received one response from the insurance domain. The respondents in the \textit{other} category are those who previously worked in software companies and worked, at the time of the survey, in academia.   
Each respondent provided a rating for each of the 46 statements (25x46) and six explanations for the three statements in most positive and most negative ratings, respectively (25x6), resulting in 1300 data points.

The demographic information of the participants (their role, and experience in development, and code review) is provided in Table~\ref{Tab:Demo}. The respondents have varying experiences from two to 30 years and work in 10 different roles such as developer, architect, and tester. Moreover, 60\% of the participants have a Master's degree, 30\% a Bachelor's degree, and 10\% have a Ph.D. degree. The respondents who provided their company names covered four different domains and seven different large companies. The details that could trace back to the respondents, such as their names and company names, are not provided to ensure the confidentiality of our respondents.

\begin{table}[t]
\centering
\scriptsize
\caption{Demographic information of the survey participants}
\label{Tab:Demo}
\begin{tabular}{llrr|llrr} 
\toprule
 \textbf{ID} & \textbf{Role} & \multicolumn{2}{c|}{\textbf{Experience (years)}} & \textbf{ID} & \textbf{Role} & \multicolumn{2}{c}{\textbf{Experience (years)}} \\
 & & \textbf{Development} & \textbf{Code review} & & & \textbf{Development} & \textbf{Code review} \\ 
\midrule
P1 & Expert engineer & 15 & 10 & P2 & Testing & 6 & 5 \\ 
P3 & Software engineer & 4 & 4 & P4 & Quality & 3 & 3 \\ 
P5 & Developer & 5 & 2 & P6 & Developer & 8 & 6 \\ 
P7 & Developer & 3 & 3 & P8 & Software engineer & 12 & 12 \\ 
P9 & Developer & 7 & 7 & P10 & Developer & 6 & 3 \\ 
P11 & Manager & 14 & 10 & P12 & Architect & 6 & 4 \\ 
P13 & Quality & 20 & 15 & P14 & Architect & 25 & 10 \\ 
P15 & Testing & 10 & 3 & P16 & Developer & 14 & 14 \\ 
P17 & Developer & 15 & 7 & P18 & Developer & 5 & 3 \\ 
P19 & Developer & 15 & 15 & P20 & Designer & 16 & 12 \\ 
P21 & Software engineer & 30 & 30 & P22 & Software engineer & 5 & 5  \\
P23 & Automation Engineer & 6 & 2 & P24 & Architect & 8 & 8  \\
P25 & Developer & 10 & 2 & \\
\bottomrule
\end{tabular}

\end{table}

\subsection{Agreement level - Rating of statements}\label{subsec:agreementlevel}
\begin{figure*}[!h]
    \centering
   \scalebox{0.65}{
    \includegraphics{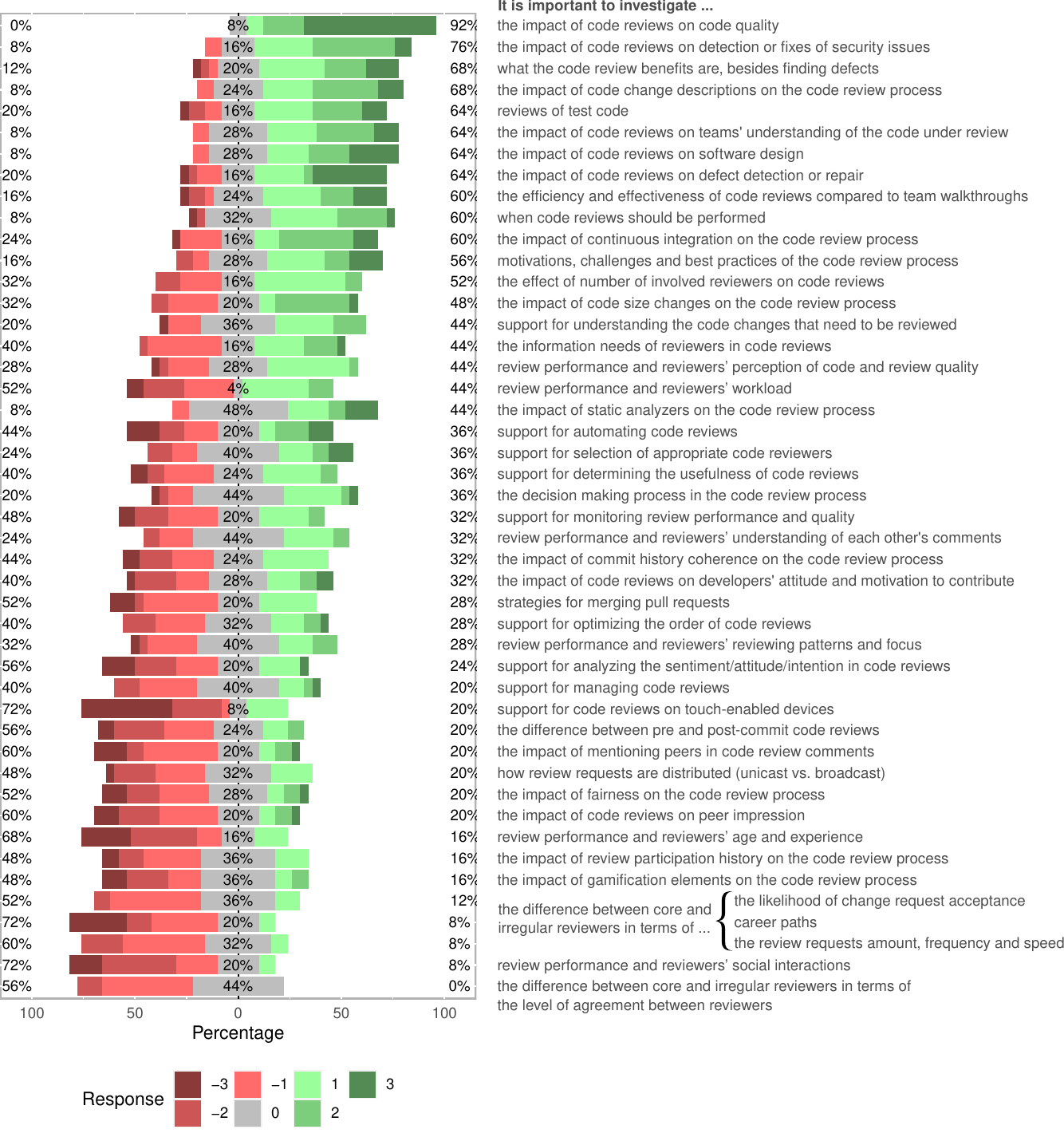} 
    }
    \caption{Agreement level of the survey respondents. \small{\textbf{Note}: All statements start with a prefix mentioned at the top of all statements.}}
    \label{fig:SurveyRes}
   
\end{figure*}

The agreement levels of 25 participants is illustrated in Figure~\ref{fig:SurveyRes}. The vertical axis shows the 46 statements, while the horizontal axis shows the percentage on the agreement levels. The statements are sorted from most to least agreement. 

Three out of the top five statements are related to the impact \emph{of} and \emph{on} code reviews. In addition, the benefits of code/test reviews have high positive ratings; 92\% of the respondents agreed on '\emph{It is important to investigate the impact of code reviews on code quality'}. Similarly, four of the last five statements are on investigating the difference between core and irregular reviewers. None of the participants agreed with Statement '\emph{It is important to investigate the difference between core vs irregular reviewers in terms of the level of agreement between reviewers'}. The statement '\emph{It is important to investigate support for code reviews on touch-enabled devices'} received the most negative ratings. However, we can see from Figure~\ref{fig:SurveyRes} that there is no consensus on most of the statements. In other words, most statements received both positive and negative as well as neutral responses. In some cases (35\% of the statements) the difference in positive and negative ratings is not vast (less than 20\%). For example, '\emph{It is important to investigate support for determining the usefulness of code reviews'} has 40\% negative, and 36\% positive responses which is a difference of only 4\%. A deeper look at the differences is needed.

\begin{figure}
\centering
\begin{subfigure}[h]{.5\textwidth}
  \centering
  \includegraphics[width=.9\linewidth]{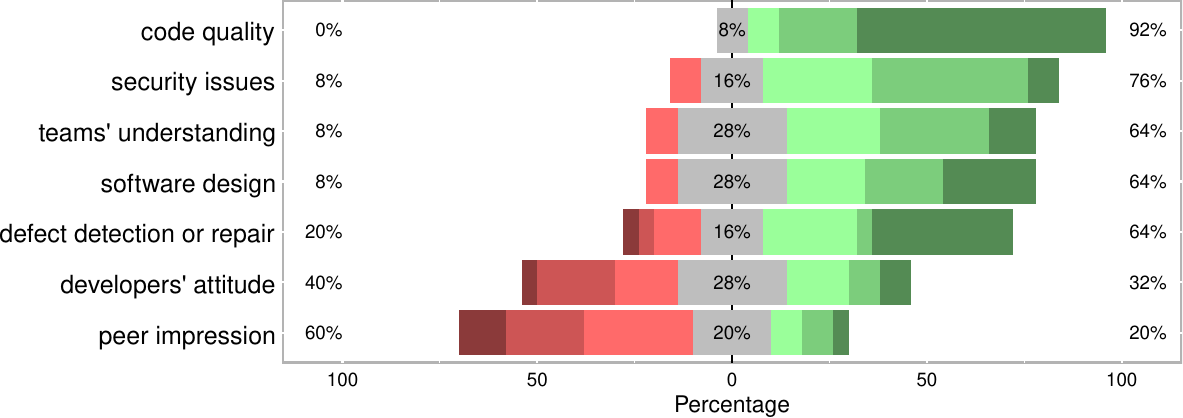}
  \caption{Impact of code reviews on product quality and human aspects (IOF)}
  \label{fig:SurveyIOF}
\end{subfigure}%
\hfill
\begin{subfigure}[h]{.5\textwidth}
  \centering
  \includegraphics[width=.9\linewidth]{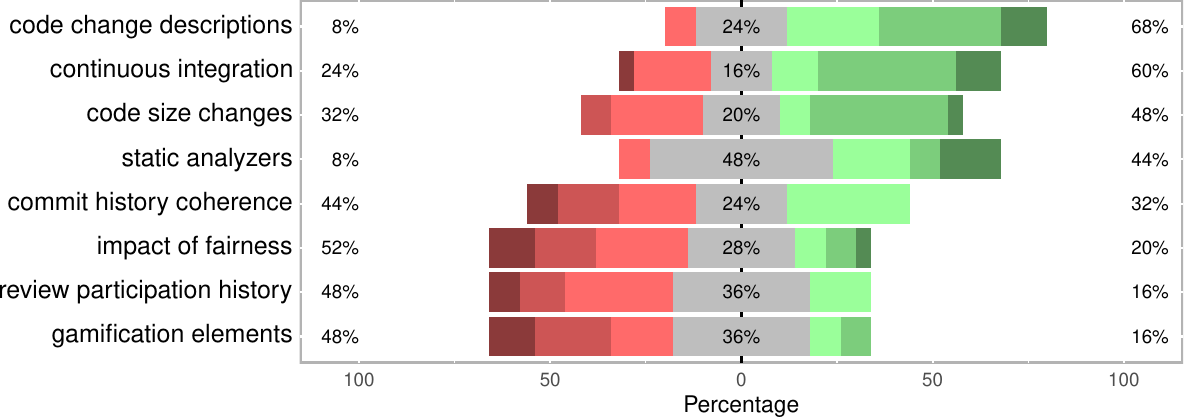} 
  \caption{Impact of software development processes, patch characteristics,
and tools on modern code reviews (ION)}
  \label{fig:SurveyION}
\end{subfigure}
\begin{subfigure}[h]{.5\textwidth}
  \centering
  \includegraphics[width=.99\linewidth]{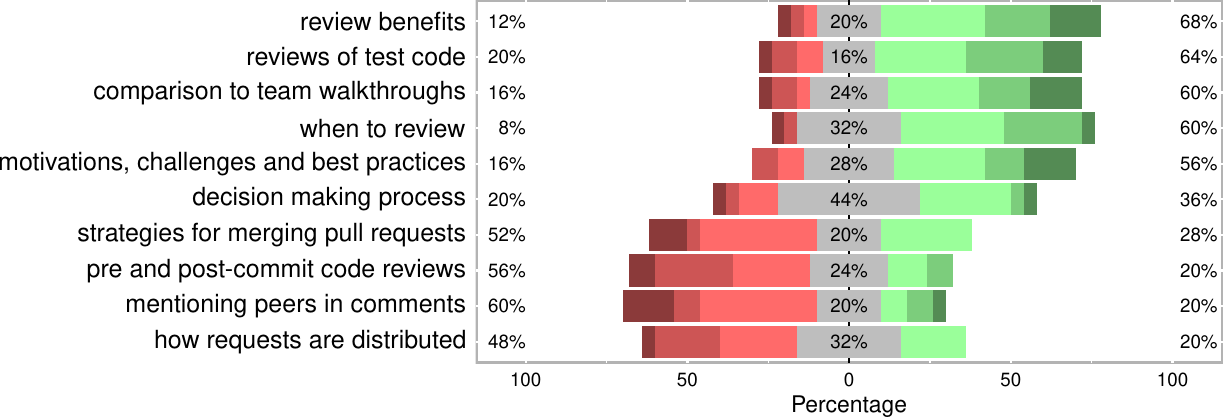} 
   \caption{Modern code review process properties (CRP)}
    \label{fig:SurveyCRP}
\end{subfigure}%
\hfill
\begin{subfigure}{.5\textwidth}
  \centering
  \includegraphics[width=.99\linewidth]{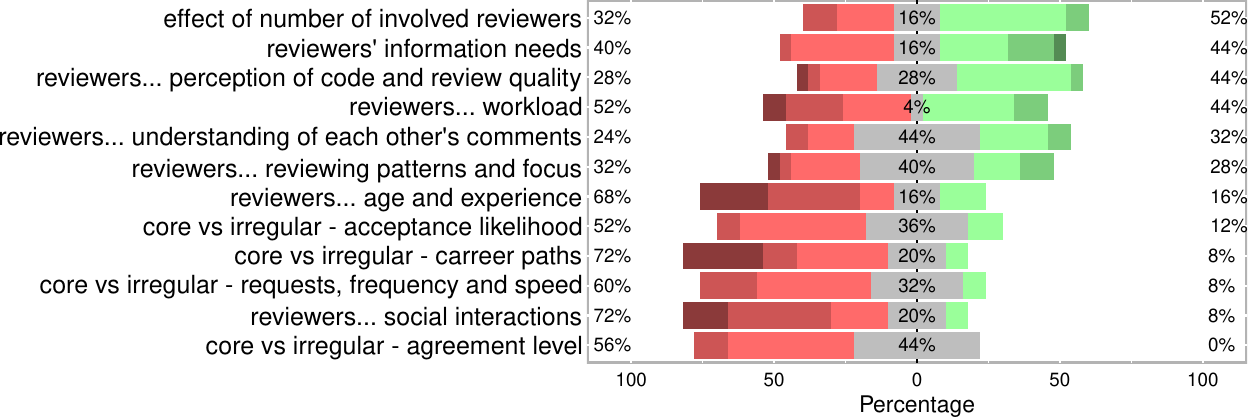} 
    \caption{Human and organizational factors (HOF)}
    \label{fig:SurveyHOF}
\end{subfigure}
\begin{subfigure}[h]{.5\textwidth}
    \centering
    \includegraphics[width=0.99\linewidth]{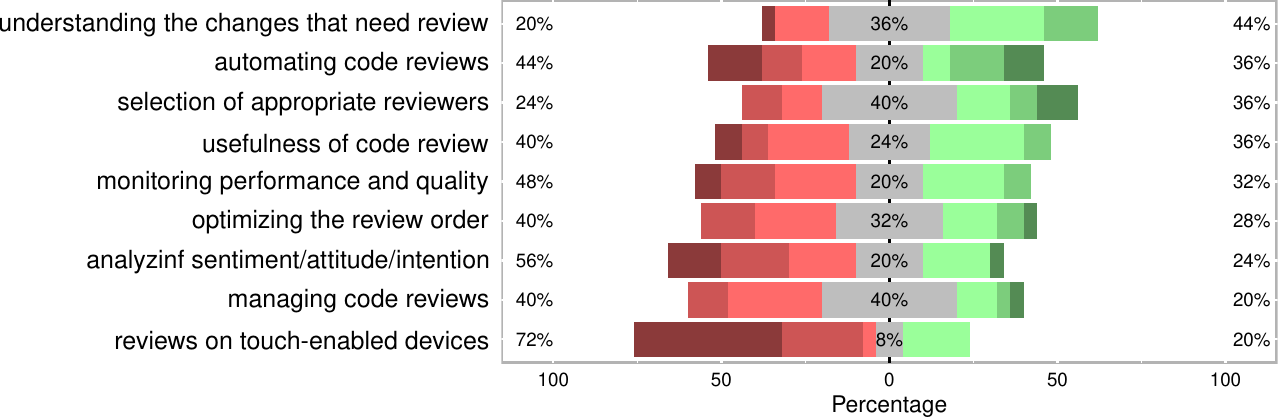} 
    \caption{Support systems for code reviews (SS)}
    \label{fig:SurveySS}
\end{subfigure}
\caption{Ratings for the five themes}\label{fig:ratings}
\vspace{-4mm}
\end{figure}

We grouped the rating on the five themes to see how the ratings vary within each. As seen in Figure~\ref{fig:ratings}, the theme ``Impact of code reviews on product quality and human aspects'' (IOF, see Figure~\ref{fig:SurveyIOF}) has a received the most positive response. However, within the theme, the impact of code reviews on product quality (i.e., code quality, security issues, software design and defect detection or repair) received a more positive response compared to human aspects, particularly on developers' attitude and peer impression. This indicates that practitioners perceive that research on the impact of code reviews on the outcome (e.g.,quality) is more important than research on human aspects. This observation is further corroborated by the fact that the theme ``Human and organizational factors'' (HOF) had the second least agreement, short of the theme ``Support systems for code reviews'' (SS).

Figure~\ref{fig:SurveyION} depicts the ratings on the theme ``Impact of software development processes, patch characteristics, and tools on modern code reviews'' (ION). More than 50\% of the respondents perceive the investigation of code change description, continuous integration, code size changes, and static analysers on code review process to be important. On the other hand, the impact of commit history coherence, fairness, review participation history, and gamification on code reviews are not considered as important. We observe as well in the ION theme that practitioners are more negative towards research on human aspects such as impact of fairness in code reviews, which received the highest negative rating in the ION theme.


The investigation in the theme ``Modern code review process properties'' (CRP) is considered important, especially research on the investigation of benefits, challenges, and best practices, (see Figure~\ref{fig:SurveyCRP}). However, some of the topics such as the process for distributing review requests, and merging pull requests were not considered as important. 

In the theme ``Human and organizational factors'' (HOF), research on the effect of the number of involved reviewers, reviewers' information needs, reviewers perception of code and review quality, and reviewers understanding each others' comments was by the majority perceived as important. However, 72\% of the respondents did not agree on the need to investigate reviewers' career paths and social interactions, as seen in Figure~\ref{fig:SurveyHOF}. In addition, 68\% of the respondents did not perceive research on the reviewers' age and experience to be important. 

In the theme ``Support systems for code reviews" (SS), only research on support for understanding what changes need review and selection of appropriate reviewers was perceived as important, as shown in Figure~\ref{fig:SurveySS}. Support for code reviews on touch-enabled device received most negative response where 72\% of the respondents gave negative ratings. It is rather surprising that this theme received the least agreement overall, given that it is the theme with the majority of publications. 

When looking at the statements grouped in themes, there is a clear trend for the practitioners' preference on research that investigates causal relationships between code reviews and factors relevant for software engineering in general (themes ION and IOF). There is also a strong interest on modern code review process properties. Surprisingly, research on human and organizational factors as well as support systems for code reviews was not perceived as important by practitioners, which together represent nearly 70\% (164 out of 244) of the primary studies from our mapping study.

\subsection{Factor analysis}\label{Sec:differentViews}
We further analyzed the survey data to identify patterns in the respondents' viewpoints using factor analysis, as suggested by the Q-Methodology. 
In the survey, we asked respondents to put only a fixed number of statements per rating for example, only 3 statements in each of the -3 and 3 ratings. However, due to an error in the survey tool, four respondents could put more than the desired statements in some ratings. Therefore, we only included 21 out of 25 valid participants responses in Q-method analysis. As mentioned in Section~\ref{Qmethod}, the participants rate the statements which is represented in a Q-Sort. For example a Q-Sort of one participation for 46 statements is as follows
(-3	3	2	3	2	-3	2	0	-3	2	0	0	-2	-2	0	2	-2	1	-1	3	0	1	-1	1	-1	1	0	0	-1	1	-1	-1	0	-1	1	-1	1	1	1	0	0	-2	0	-2	-1	0), where each value is the rating given by the participant for the statement. The Q-Sorts of all participants is used as input for factor analysis. 

\begin{figure}[h]
    \centering
    \includegraphics[width=\textwidth/2]{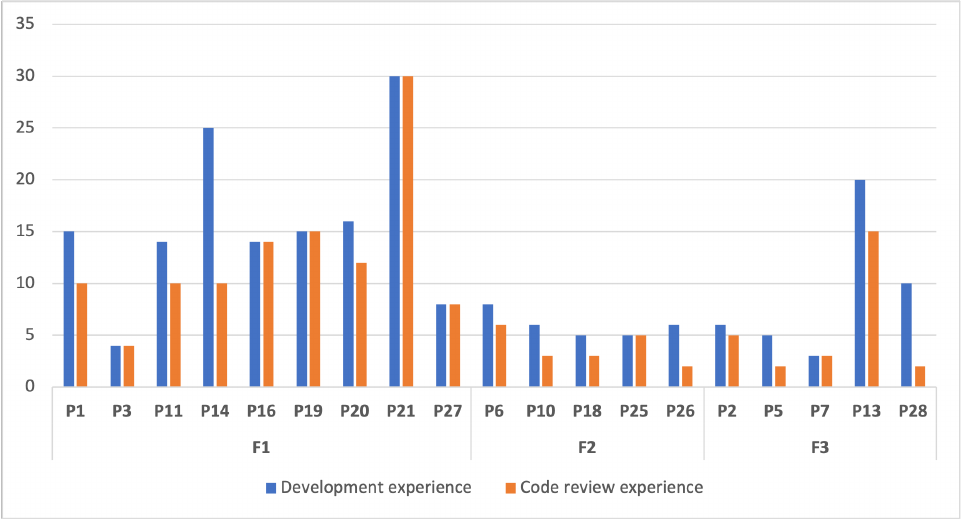}
    \caption{Participants in each factor}
    \label{fig:participant}
    \vspace{-4mm}
\end{figure}

The steps followed in the Q-method analysis are (see the result of each intermediate step in the Q-method report available online~\cite{deepika_badampudi_2022_7066821}):
\begin{enumerate}
\item Creating an initial matrix - An initial two-dimensional matrix is created (statements x participants), where the value of each cell is the rating given by the participants (between -3 to 3).
\item Creating correlation matrix - A correlation matrix between each Q-Sort (i.e., participant ratings) is generated using Pearson correlation coefficient test. 
\item Extracting factors and creating factor matrix - New Q-Sorts called factors are extracted which are the weighted average Q-Sorts of all participants with similar ratings.  A factor represents Q-Sort of a hypothetical participant representing a similar viewpoint. We used principal component analysis (PCA) to extract the factors. The two-dimensional factor matrix is created (participants x factors). The value of each matrix cell is the correlation between the participants Q-Sort and the factors called factor loading. A higher loading value indicates more similarity between the participant and the factor.  
\item Calculating rotated factor loading - To clarify the relation among factors and increase explanatory capacity of the factors resulting from PCA, we conducted varimax factor rotation. Only a few factors are selected that represent the maximum variance. We used both a quantitative and qualitative approach to find select the number of factors. The quantitative criteria recommend in the literature are as follows \cite{cartaxo2019esem}: a minimum of two loading Q-Sorts are highly correlated to the factor; (2) the composite reliability is greater or equal to 0.8 for each factor; (3) eigenvalues are above 1 for each factor, and (4) the sum of explained variance percentage of all the selected factors should be between 40\% and 60\%.

\begin{table}[h]
\centering

\scriptsize
\caption{Factor characteristics from Q-methodology}
\label{tab:FactorChar}
\begin{tabular}{llll} 
\toprule
\textbf{Characteristics}  & \textbf{F1} & \textbf{F2} & \textbf{F3}           \\
\midrule
Average reliability coefficient  & 0.80        & 0.80        & 0.80         \\
Number of loading Q-sorts        & 9.00        & 5.00        & 5.00         \\
Eigenvalues                      & 3.53        & 3.24        & 3.17         \\
Percentage of explained variance & 16.81       & 15.43       & 15.07        \\
Composite reliability            & 0.97        & 0.95        & 0.95         \\
Standard error of factor scores  & 0.16        & 0.22        & 0.22         \\
\bottomrule
\end{tabular}
\end{table}
As shown in Table~\ref{tab:FactorChar}, when we select three factors all of the above criteria are satisfied. We also performed a qualitative analysis and excluded solutions with more than three factors as there were few or no distinguishable statements. For example, a statement is distinguishable when its rank in one factor differs from all other factors. 
\item Finalising factor loading - The rotated factor loadings from the previous step are finalised by flagging the Q-Sorts that best represent the factors. We flagged the Q-Sorts based on the follow criteria: (1) Q-Sorts with factor loading higher than the threshold for p-value < 0.05; (2) Q-Sorts with square loading higher than the sum of square loadings of the same Q-Sort in all other factors. As seen in Table~\ref{tab:FactorChar}, the sum of number of loading Q-Sorts is 19 which means two respondents could not be significantly loaded into any of the factors.  
\item Calculating the z-scores and factor scores - The z-scores and factor scores indicate the statement's relative position within the factor. The z-score is a weighted average of the values of flagged participants' ratings given to a statement in the factor. Factor scores are based on ordering z-scores and mapping to the Q-Sort structure (-3 to 3); they are integer values instead of continuous. Factor scores are important for factor interpretation. 
\item Identifying distinguish statements - As mentioned in Step 4, a statement is distinguishable when its rank in one factor differs from all other factors. The factor scores from Step 5 are used to identify distinguished statements that represent the factor and used for factor interpretation. If there is a significant difference (more than 0.05) in factor score of a statement in one factor from all other factors that the statement is identified as distinguished statement. The distinguished statements in Factor 1 is provided in Table~\ref{tab:Statement-SurveyDistF1}, Factor 2 in Table~\ref{tab:Statement-SurveyDistF2}, and Factor 3 in Table~\ref{tab:Statement-SurveyDistF3}. 
\end{enumerate}
Figure~\ref{fig:participant} provides a summary of the respondents' experience in development and code reviewing in each factor. It is clear that respondents with more experience are grouped in Factor 1 compared to other factors. We provide an interpretation of the factors in the next subsections. 

\subsubsection{Factor 1 interpretation - It is important to investigate the code reviewer as a subject}\label{Sec:Factor1}
 
\begin{table}[h]
\centering
\scriptsize
\caption{Distinguishing statements in Factor 1}
\label{tab:Statement-SurveyDistF1}
\begin{tabular}{p{0.8\textwidth}rrr} 
\toprule
\textbf{Statements (It is important to investigate \ldots)} & \textbf{F1} & \textbf{F2} & \textbf{F3}  \\ 
\midrule
IOF: \ldots the impact of code reviews on teams' understanding of the code under review & \textcolor[rgb]{0,0.502,0}{3} & 1 & 1\\ 
IOF: \ldots the impact of code reviews on developers' attitude and motivation to contribute & \textcolor[rgb]{0,0.502,0}{1} & -1 & -1 \\ 
IOF: \ldots the impact of code reviews on peer impression in terms of trust, reliability, perception of expertise, and friendship & \textcolor[rgb]{0,0.502,0}{0} & -3 & -2 \\ 
HOF: \ldots the difference between core and irregular reviewers in terms of career paths & -1 & -3 & -3\\ 
SS: \ldots support for understanding the code changes that need to be reviewed & \textcolor{red}{-1} & 1 & 0\\
\bottomrule
\end{tabular}
\end{table}

Table~\ref{tab:Statement-SurveyDistF1} shows the distinguishing statements on Factor 1 which represents 43\% of the respondents and explains 16.81\% of the variance in responses. As seen in Figure~\ref{fig:participant}, participants loaded in Factor 1 have more experience. They have expert/senior roles in architecture and design, an average of 16 years experience in software development, and 13 years of code review experience.  
Participants loaded in Factor 1 are more positive regarding the impact of code review on human factors than the ones loaded in Factors 2 and 3. For example, statements regarding the teams' understanding of the code under review, developers' attitude, and peer impression are perceived to be important in Factor 1. Regarding the impact of code reviews on teams' understanding, one of the respondents in Factor 1 wrote: \emph{P27: "Without understanding the requirement of the code, there is no point to review the code"}. Another respondent was interested in the investigation of knowledge sharing, he wrote \emph{P3: "Code reviews enable knowledge sharing"}. Research on the impact of code review on developers' attitude is considered important as considerable amount of effort goes in reviewing code. A participant wrote: \emph{P14: "Everyone needs to see the importance of better quality"}. However, respondents in Factors 2 and 3 disagree. One of the respondents in Factor 2 wrote: \emph{P26: "this is more of an individual's approach towards any work. Once a reviewer is made to follow the correct set of principles, this [investigation on developers' attitude] can be eliminated"}. 

All respondents display a neutral or even negative attitude towards the importance of investigating the impact of peer impression on code reviews. They feel that people should be objective and not be influenced by peer impressions. One of the respondents in Factor 1 wrote: \emph{"P1: It should not be necessary to do research on the obvious fact that people should be responsible"}. Similarly, respondents in Factor 1 are less negative compared to F2 and F3 about the importance to investigate the difference between core and irregular reviewers in terms of their career paths. On the other hand, respondents in Factor 1 are more negative compared to F2 and F3 about the importance to investigate support for understanding the code changes that need review. One of the respondents wrote: \emph{"P14: Everything should be reviewed, this is a non-question"}. However, the practitioner interpreted the question as "understanding what should be reviewed" rather than understanding the code under review. Despite the potential misinterpretation, this statement has been ranked as most important statement in the solutions theme (see Figure~\ref{fig:SurveySS}). 

\subsubsection{Factor 2 interpretation - It is not important to investigate human aspects related to code review}\label{Sec:Factor2} 

\begin{table}
\centering
\scriptsize
\caption{Distinguishing statements in Factor 2}
\label{tab:Statement-SurveyDistF2}
\begin{tabular}{p{0.8\textwidth}rrr} 
\toprule
\textbf{Statements (It is important to investigate \ldots)} & \textbf{F1} & \textbf{F2} & \textbf{F3}  \\ 
\midrule
IOF: \ldots the impact of code reviews on defect detection or repair                        & 1                                & \textcolor[rgb]{0,0.502,0}{3}    & 0                                 \\ 

ION: \ldots the impact of continuous integration on the code review process                            & 0                                & \textcolor[rgb]{0,0.502,0}{3}    & 0                                 \\ 

CRP: \ldots the impact of mentioning peers in code review comments                      & -2                               & 0                                & -1                                \\ 

HOF: \ldots review performance and reviewers’ age and experience                     & -2                               & \textcolor{red}{-3}              & -1                                \\ 

HOF: \ldots review performance and reviewers’ understanding of each other's comments & 0                                & \textcolor{red}{-1}              & 0                                 \\ 

HOF: \ldots review performance and reviewers’ reviewing patterns and focus           & 0                                & \textcolor{red}{-1}              & 1                                 \\ 

HOF: \ldots review performance and reviewers’ perception of code and review quality  & 1                                & \textcolor{red}{-1}              & 1                                 \\ 

HOF: \ldots the difference between core and irregular reviewers in terms of the likelihood of change request acceptance         & 0                                & \textcolor{red}{-2}              & 0                                 \\ 

HOF: \ldots the difference between core and irregular reviewers in terms of the level of agreement between reviewers               & 0                                & \textcolor{red}{-2}              & 0                                 \\
\bottomrule
\end{tabular}
\end{table}

Table~\ref{tab:Statement-SurveyDistF2} shows the distinguishing statements on Factor 2, which represents 24\% of the participants and explains 15.43\% of the variance in responses. The respondents grouped in this factor have less experience compared to the respondents in Factor 1 and 3 (see Figure~\ref{fig:participant}). They have roles in development and testing with an average of 6 years experience in software development and 4 years of code review experience. 
Respondents in this factor are more positive about research on the impact of code review on defect detection or repair and impact of continuous integration than research on human factors. On the importance of defect detection or repair, one of the respondents wrote: \emph{P18: "This [defect detection] is generally why code reviews take place - it is interesting to perform a more formal causal analysis on this [the impact of code reviews on defect detection]"}. 

Respondents do not see the importance of investigating human aspects unlike in Factor 1, where respondents with more experience are positive towards investigations on human factors. In this factor, more importance is given to having good code review guidelines as stated by one of the participants: \emph{P25: "Standard review procedure should be independent of individual/team members' age and experience}". 

\subsubsection{Factor 3 interpretation - It is more important to investigate the support for optimizing code reviews than support for analyzing human aspects.}\label{Sec:Factor3}

\begin{table}
\centering
\scriptsize
\caption{Distinguishing statements in Factor 3}
\label{tab:Statement-SurveyDistF3}
\begin{tabular}{p{0.8\textwidth}rrr} 
\toprule
\textbf{Statement (It is important to investigate \ldots)}                          & \textbf{F1} & \textbf{F2} & \textbf{F3}  \\ 
\midrule
IOF: \ldots the impact of code reviews on software design                        & 3                                & 2                                & 0                                 \\ 
ION: \ldots the impact of code change descriptions on the code review process               & 2                                & 2                                & 0                                 \\ 
ION: \ldots the impact of code size changes on the code review process                      & 1                                & 1                                & \textcolor{red}{-1}               \\ 
CRP: \ldots when code reviews should be performed                         & 2                                & 2                                & 1                                 \\ 
SS: \ldots support for analyzing the sentiment/attitude/intention in code reviews & 0                                & -1                               & \textcolor{red}{-3}               \\ 
SS: \ldots support for optimizing the order of code reviews             & -1                               & -1                               & \textcolor[rgb]{0,0.502,0}{1}     \\
\bottomrule
\end{tabular}
\end{table}

Table~\ref{tab:Statement-SurveyDistF3} shows the distinguishing statements on Factor 3, which represents 24\% of the participants and explains 15.07\% of the variance in responses. Respondents in this factor have mainly testing roles and an average of 9 years experience in software development and 5 years of code review experience.

Overall respondents in Factor 3 are less positive about research on the impact on and of code reviews and code review process. They are more interested in research on the support for optimizing code reviews than analyzing human aspects. 

We did not get any explanations for the ratings as most of the ratings are between -2 to 2. For the -3 rating the respondent had no comments.  

\section{Comparing the state-of-the-art and the practitioners' perceptions}\label{sec:comp}
In this section, we answer RQ3 --- \emph{To what degree are researchers and practitioners aligned on the goals of MCR research?} --- by juxtaposing the results from the mapping study (Section~\ref{Sec:msresults}) and the responses from the survey (Section~\ref{sec:surveyresults}).

\subsection{Comparing the number of research articles and the practitioners' perceptions}\label{sec:compNo.ofArticles}
\begin{figure}[t]
    \centering
    \includegraphics[width=0.6\textwidth]{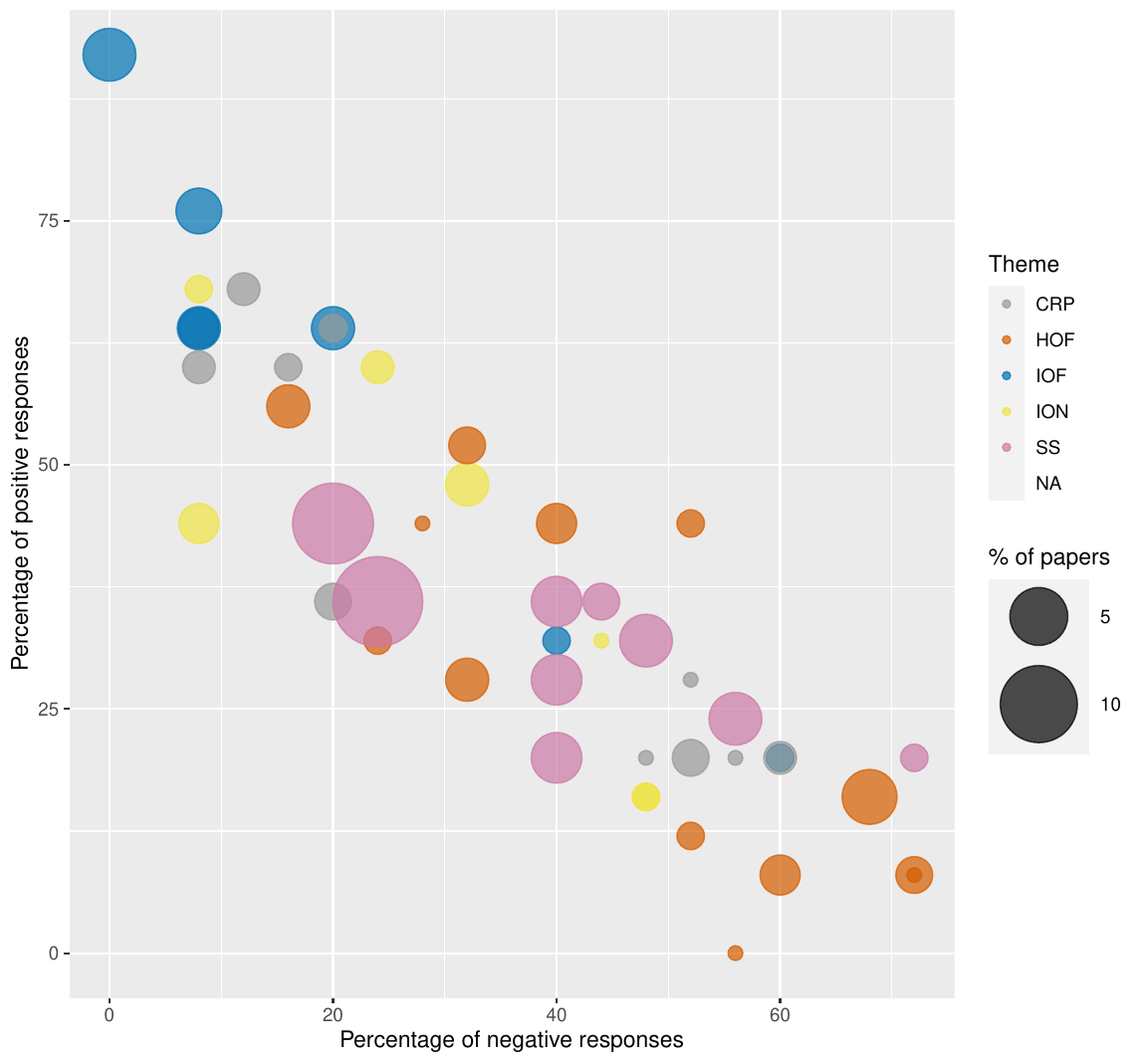} 
    \caption{Comparison of literature and practitioners' perceptions}
    \label{fig:MappingSurveyLR}
\end{figure}

In Figure~\ref{fig:MappingSurveyLR}, we map survey responses, the percentage of papers representing a survey statement, and modern code review themes. The percentage of negative and positive responses for each statement is shown on the x- and y-axes, respectively. Each bubble represents a statement from the survey and its size indicates the percentage of representing papers. The different colors represent the five themes we identified in the mapping study. 

In addition, we evaluated if there is a statistical correlation between the number of papers and practitioners' perceptions. Using Shapiro-Wilk normality test we determined that our data is normally distributed. We then conducted a Pearson correlation test to evaluate if there is a significant relation between the ratings and the number of papers in different themes. The result of the correlation test is provided in Table~\ref{tab:ThemePapersCorrelation}, the statistically significant results are bold.

Figure~\ref{fig:MappingSurveyLR} accentuates a result we reported on the agreement levels in Section~\ref{subsec:agreementlevel}: while there is considerable research on solution support (SS), and human and organizational factors (HOF), as indicated by the number and size of bubbles, practitioners seem to have a rather negative attitude towards the research done in this theme. None of the solution statements received more than 50\% positive responses. Within this theme, research on support for understanding the code changes that need to be reviewed and support for the selection of appropriate reviewers received the most positive responses and was also associated with the most papers. This is a good example of alignment between research and practitioners' interest. The positive alignment is also confirmed in the correlation test as the solutions that have fewer publications received also more negative ratings (c.f., Table~\ref{tab:ThemePapersCorrelation}).  On the topic of reviewer selection, one of the respondents noted that \emph{"P9: The most effective review is the one done by developers who are the most familiar with a particular functionality or have worked on a similar functionality on a different project. I think there is no helping tool to tell who is the most appropriate reviewer.}" Several studies propose or evaluate tools that do just exactly that. While the respondents' answer is certainly not representative, more focus on knowledge translation and transfer to practitioners about existing solutions can be a beneficial target for researchers in this area. Furthermore, as seen in Section \ref{Sec:Solutions}, only two out of 36 solutions supporting the reviewer recommendation provide links to the tools, which could explain why practitioners are unaware of the existing solutions. 

Looking at Figure~\ref{fig:MappingSurveyLR}, we see more negative than positive responses for statements related to human and organization factors (HOF). However, we did not find any statistical significant relation between the number of papers in the HOF theme and the rating, as indicated in Table~\ref{tab:ThemePapersCorrelation}. The most positively received statement is related to investigating the effect of the number of involved reviewers in code reviews. The statement investigating review performance and reviewers' age and experience in this theme is associated with the most studies but it is also perceived mostly negatively. For example, a respondent wrote: \emph{P2: "Age and experience is less important than code knowledge or ability to read code. An 18 year old with no experience writes the best comments, then that is the person I will invite to review"}. Another participant elaborated more on the age factor: \emph{P7: "I don't understand how the age of reviewer can help in performance, Experience to certain extent but that doesn't mean the experienced person knows new technologies that are emerging so this statement should be viewed as 2 separate things with respect to experience yes important to investigate to certain extent. But with respect to age some younger ones are actually doing more reviews now a days"}. Another respondent emphasised the importance on a standard review process being more important than reviewer age and experience: \emph{P25: "Standard review procedure is to be independent of individual/team members' age and experience"}. 

Looking at the top-left corner of Figure~\ref{fig:MappingSurveyLR}, the area with high positive and low negative ratings is dominated by statements related to research on the impact of code reviews on product quality and human aspects (IOF) and modern code review process properties (CRP). Although, we can see that only the relation between the ratings and papers in the IOF theme is statistically significant (c.f., Table~\ref{tab:ThemePapersCorrelation}). This result indicates that practitioners are interested in research that investigates causal relationships, as indicated by a respondent \emph{P11: "Understanding how people approach and make decisions when performing a code review may open up some other interesting questions in how to structure and format code reviews to be more effective"}. However, there is only a relatively low number of studies in this area.

\begin{table}[]
\caption{Correlation between the number of papers published within a theme and practitioners' perceptions}
\label{tab:ThemePapersCorrelation}
\resizebox{\linewidth}{!}{%
\begin{tabular}{lllllllllll}
\hline
\multicolumn{1}{c}{\multirow{2}{*}{Pearson correlation}}                          & \multicolumn{2}{c}{IOF}                             & \multicolumn{2}{c}{ION}                             & \multicolumn{2}{c}{CRP}                             & \multicolumn{2}{c}{HOF}                             & \multicolumn{2}{c}{SS}                              \\
\multicolumn{1}{c}{}                                                              & \multicolumn{1}{c}{p value} & \multicolumn{1}{c}{r} & \multicolumn{1}{c}{p value} & \multicolumn{1}{c}{r} & \multicolumn{1}{c}{p value} & \multicolumn{1}{c}{r} & \multicolumn{1}{c}{p value} & \multicolumn{1}{c}{r} & \multicolumn{1}{c}{p value} & \multicolumn{1}{c}{r} \\
\midrule
\begin{tabular}[c]{@{}l@{}}Positive rating - \\ Percentage of papers\end{tabular} & \textbf{0.06555}            & \textbf{0.9693013}    & 0.4908                      & 0.2869428             & 0.2859                      & 0.400119              & 0.3291                      & -0.2942445            & 0.09438                     & 0.5901029             \\
\begin{tabular}[c]{@{}l@{}}Negative rating - \\ Percentage of papers\end{tabular} & \textbf{0.01202}            & \textbf{-0.8646518}   & 0.3612                      & -0.3741424            & 0.125                       & -0.55                 & 0.2675                      & 0.3321971             & \textbf{0.009847}           & \textbf{-0.7986266}   \\ \hline
\end{tabular}
}
\end{table}

\subsection{Comparing research impact and practitioners' perceptions}\label{sec:compImpact}
We retrieved citations of all primary studies as of August 2022. Peer citation is one way of assessing the research impact and the activity of a theme. We compared the research impact with the practitioners responses from the survey. As we have the practitioners responses on each statement, we calculated the research impact for each statement by considering the sum of citations of all primary studies representing a statement (see Table~\ref{tab:statementsfreq}). 
\begin{figure}
    \centering
    \includegraphics[width=0.8\textwidth]{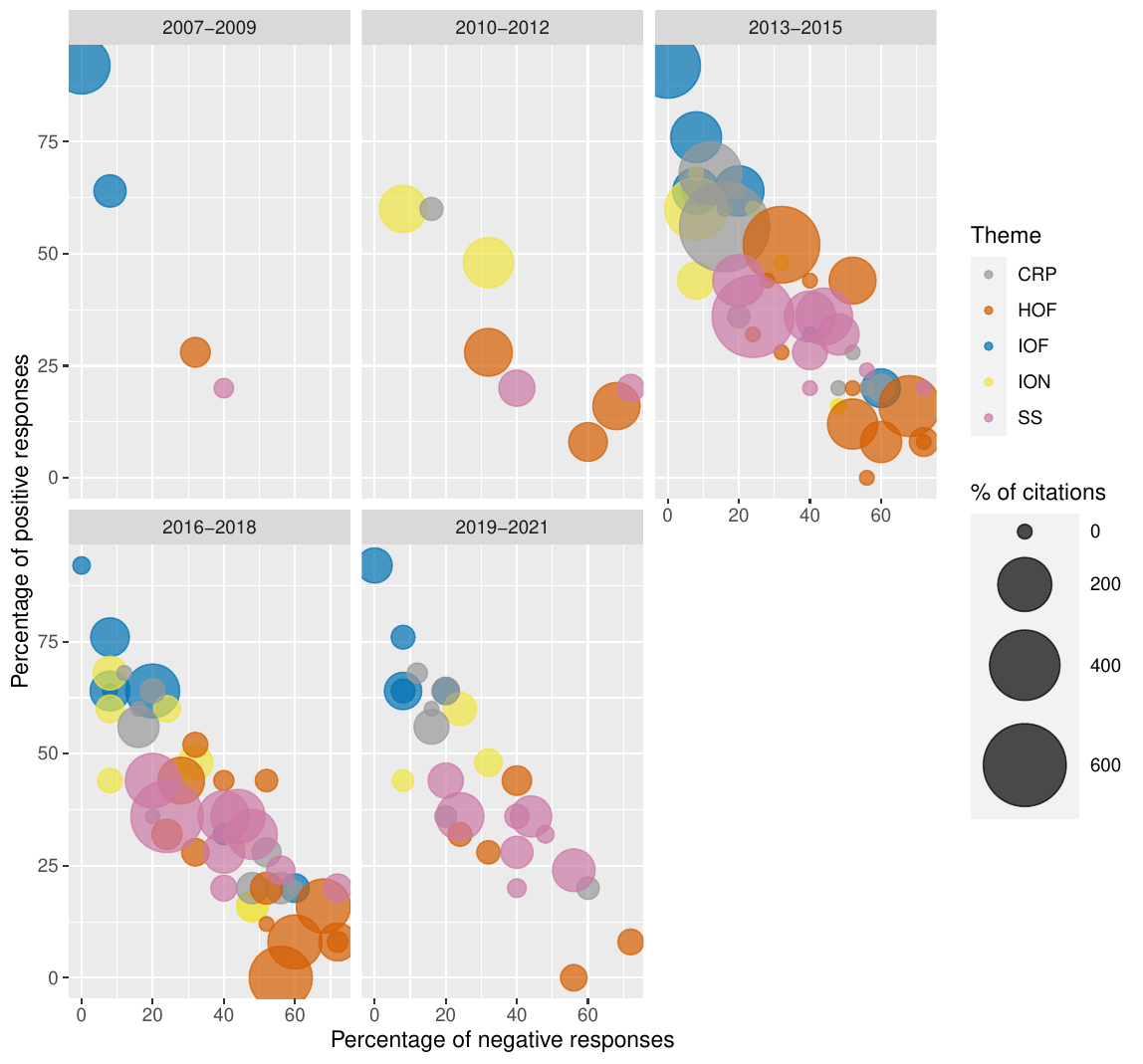}
    \caption{Comparison of research impact and practitioners’ perceptions}
    \label{fig:MappingSurveyLRImpact}
    \vspace{-4mm}
\end{figure}
\begin{table}[!htb]
\centering
\caption{Correlation between the ratings and research impact}
\label{Tab:Correlation}
\resizebox{\linewidth}{!}{%
\begin{tabular}{cllllllllll} 
\toprule
\multirow{2}{*}{Pearson correlation}                       & \multicolumn{2}{c}{2007-2009}                       &  \multicolumn{2}{c}{2010-2012}                       & \multicolumn{2}{c}{2013-2015}                       & \multicolumn{2}{c}{2016-2018}                       & \multicolumn{2}{c}{2019-2021}                             \\
                                                & \multicolumn{1}{c}{p value} & \multicolumn{1}{c}{r} & \multicolumn{1}{c}{p value} & \multicolumn{1}{c}{r} &  \multicolumn{1}{c}{p value} & \multicolumn{1}{c}{r} & \multicolumn{1}{c}{p value} & \multicolumn{1}{c}{r} & \multicolumn{1}{c}{p value} & \multicolumn{1}{c}{r}  \\
                                                \midrule
\multicolumn{1}{l}{Positive rating - Citations} & 0.1186                     & 0.8814008                               & 0.8205                      & 0.09631116             & \textbf{0.01884}                      & \textbf{0.3450907 }             & 0.2793                      & -0.1629392            & 0.9237                      & 0.01934015 ~             \\
\multicolumn{1}{l}{Negative rating - Citations} & 0.2149                       & -0.7850902                           & 0.6684                      & -0.1807715            & \textbf{0.0323}                      & \textbf{-0.3161727}            & 0.6639                      & 0.06581944          & 0.9631                     & 0.009334502             \\
\bottomrule
\end{tabular}
}
\end{table}

\begin{table}[!htb]
\caption{Correlation between the ratings and research impact on each theme}
\label{tab:ThemeImpactCorrelation}
\resizebox{\linewidth}{!}{%
\begin{tabular}{lllllllllll}
\hline
\multicolumn{1}{c}{\multirow{2}{*}{Pearson correlation}} & \multicolumn{2}{c}{IOF}                             & \multicolumn{2}{c}{ION}                             & \multicolumn{2}{c}{CRP}                             & \multicolumn{2}{c}{HOF}                             & \multicolumn{2}{c}{SS}                              \\
\multicolumn{1}{c}{}                                     & \multicolumn{1}{c}{p value} & \multicolumn{1}{c}{r} & \multicolumn{1}{c}{p value} & \multicolumn{1}{c}{r} & \multicolumn{1}{c}{p value} & \multicolumn{1}{c}{r} & \multicolumn{1}{c}{p value} & \multicolumn{1}{c}{r} & \multicolumn{1}{c}{p value} & \multicolumn{1}{c}{r} \\
\midrule
Positive rating - Citations                              & 0.06555            & 0.4091079    & 0.223                       & 0.2776709             & 0.3124                      & 0.2105235             & 0.9791                      & 0.004592727           & \textbf{0.001982}           & \textbf{0.5502959}    \\
Negative rating - Citations                              & 0.2306                      & -0.2733161            & 0.1797                      & -0.3044363            & 0.2766                      & -0.2263241            & 0.5221                      & 0.1119356             & \textbf{0.004827}           & \textbf{-0.5087684}   \\ \hline
\end{tabular}
}
\end{table}

We grouped the analysis by creating bins for the publication year, since more recent publications have likely less citations than older publications, which may have had simply more time for being cited. The primary studies are published between 2007 and 2021 (Figure~\ref{fig:MappingSurveyLRImpact}). The percentage of negative and positive responses for each statement is shown on the x- and y-axes and the colors represent the different themes. Each bubble represents a statement from the survey and its size indicates the total number of citations of all primary studies in each statement.

In addition, we evaluated if there is a statistical correlation between the research impact and practitioners' perceptions. Using Shapiro-Wilk normality test we determined that our data is normally distributed. Then we conducted a Pearson correlation test to evaluate if there is a significant relation between the ratings and the research impact in different years. Table~\ref{Tab:Correlation} shows the results of Pearson's correlation test for the different years. We also evaluated the correlation between the ratings and the research impact of papers in each theme (see Table~\ref{tab:ThemeImpactCorrelation}).

Although the overall positive ratings are low for the support systems for code reviews (SS) theme, the papers with high impact have higher positive rating compared to low impact papers. When considering all years together, the SS theme exhibits a significant negative correlation between negative ratings and research impact (r = -0.5087684 , p = 0.004827), indicating that when impact is high, the negative ratings are low. Similarly, the correlation between positive ratings and research impact is significant as well (r = 0.5502959 ,0.001982).
In the human and organization factor (HOF) theme we can see from Figure~\ref{fig:MappingSurveyLRImpact} that some of the statements that have high impact were perceived negatively by practitioners, particularly in the time frame between 2016-2018. However, we did not find any statistical significant relation between the ratings and statements in the HOF theme. In the theme related to impact of code reviews on product and human factors (IOF), we can see that statements that have high impact also received more positive ratings. We also observed a statistically significant correlation between the positive ratings and impact in the time frame between 2013-2015 (r = 0.7670108 
, p = 0.04419). We did not find any interesting patterns in the other themes.    


\section{Discussion}
\label{sec:discussion}
In this section, we summarize the research directions that emerged from analyzing the state-of-art and the practitioner survey. Furthermore, we illustrate that our findings align with the observations made by Davila and Nunes~\cite{davila2021systematic}, strengthening our common conclusions since our respective reviews cover a non-overlapping set of primary studies. Finally, we discuss the threats to validity associated with our research.

\subsection{MCR research directions} We propose future MCR research directions based on current trends and our reflections, both obtained through our mapping study and survey with practitioners. We anchor this discussion in the MCR process steps shown in Figure~\ref{fig:mcr}\footnote{Note that we did not identify any research path for \emph{Step 4 - Reviewer Notification}, indicating that this activity is already well understood.}. Next, we propose relevant research topics, along with research questions that still remain to be answered.

   \subsubsection{Preparation for code review} \hfill \\
    \textit{Understanding code to be reviewed -} Understanding the code was perceived as important by the survey respondents (c.f. Figure \ref{fig:SurveySS}). The solutions reported in the primary studies focus on a subset of patch characteristics that affect review readiness (such as \citeS{di2019effects, guo2019decomposing, wang2019cora}). However, is it possible to combine all patch characteristics into an overall score that can inform the submitter so they can improve the patch before sending it out for review?\\
    \textit{Review goal -} After understanding the code to be reviewed, the next step is to decide the review goal. The survey respondents are positive about investigating the impact of code review on code quality in general and, more specifically, security  (c.f., Figure \ref{fig:SurveyIOF}). Our primary studies findings indicate that most issues found in code reviews are related to a subset of code quality attributes such as evolvability~\citeS{beller2014modern, mantyla2008types}. Does that imply that only certain quality attributes can or should be evaluated with code reviews? \\
    \textit{Review scope - } Another aspect of preparing for code review is to decide which artifacts to review, i.e., the review scope. Test code is seldom reviewed and is not considered worthy of review~\citeS{Spadini:2018,spadini2019test}. However, the survey respondents consider test code review one of the most important research topics (c.f., Figure~\ref{fig:SurveyCRP}). In addition to test and production code, the popularity of third-party libraries (3pps) in software development is increasing, which leads to an important area of 3pp review. How to use risk-based assessment to scope the review target and the review goals to achieve an acceptable trade-off between effort and benefits?\\
    \textit{Optimizing review order} - Factors influencing the review order in open source (OSS) projects can differ from proprietary projects or may have different importance. For example, as mentioned in our primary studies~\citeS{fan2018early, Azeem:2020a, Saini:2021}, acceptance probability is one of the determinants in ordering OSS reviews (important to attract contributors). However, in proprietary projects, other determinants such as merge conflict probability may have more importance in determining the review order. 
    \begin{mybox}{Preparation}
    \begin{enumerate}
        \item[]\textbf{Understanding code to be reviewed}
        \item[Q1] What are the overall factors that affect review readiness?
        \item[Q2] How can code (patch) preparation be automated? 
        \item[]\textbf{Review goal} 
        \item[Q3]Which code quality attributes are better addressed in code reviews than other means (e.g., testing)?
        \item[]\textbf{Review scope}
        \item[Q4] What artifacts other than code should be reviewed, and how much importance should be given to these reviews?
        \item[Q5] How can risk assessment be used to determine when to review the security of 3pps?
        \item[] \textbf{Optimizing review order}
        \item[Q6] How do factors determining review order relate to project type (OSS/proprietary)?
    \end{enumerate}
    \end{mybox}

\subsubsection{Reviewer selection} \hfill \\
\textit{Appropriate reviewer selection -} The primary studies focus on identifying "good reviewers" based on certain predictors such as pull request content similarity~\citeS{Nafiz:2020, jiang2017should, xia2015should, ying2016earec}. However, how much do "good reviewers" differ in review performance from "bad reviewers"? \\
\textit{Number of  reviewers -} The primary studies establish a correlation between the number of reviewers and the review performance~\citeS{24, 27, Santos:2017}. However, what are the factors determining the optimal number of reviewers? 
\begin{mybox}{Reviewer selection}
    \begin{enumerate}
        \item[]\textbf{Appropriate reviewer selection} 
        \item[Q1]What is the impact of selecting non-recommended reviewers?
        \item[Q2] Does it matter to choose the highest ranked reviewer or follow the recommended review order?
        \item[]\textbf{Number of reviewers} 
        \item[Q3]What are the factors that determine the optimal number of reviewers for a given project, and what should be their responsibilities (security, test code review, or requirements review) when multiple reviewers are involved?
    \end{enumerate}
    \end{mybox}
 
\subsubsection{Code checking} \hfill \\ 
After the preparation and reviewer selection step, the reviewers are notified and the actual review takes place. It is valuable to monitor the review process to learn new insights that can be codified in guidelines. It is known that code review can identify design issues~\citeS{8, 32, han2021understanding, paixao2019impact}. How can this identification be used as an input to create or update design guidelines? In addition, the primary studies found that good reviewers exhibit different code scanning patterns than less good reviewers~\citeS{Uwano:2007, Sharif:2012, Begel:2018,Chandrika:2018}. Such findings should be used to propose/develop solutions that harvest this expertise from reviewers. 

\begin{mybox}{Code checking}
    \begin{enumerate}
    \item[]\textbf{Design support} 
        \item[Q1] How can design rules/guidelines be extracted and updated based on the design issues identified and corrected through code reviews?
        \item[]\textbf{Scanning patterns} 
        \item[Q2] How can reviewers' focus patterns, extracted by eye tracking from code reviewers, be used to model and eventually transfer reviewing expertise? 
    \end{enumerate}
    \end{mybox}
    
\subsubsection{Reviewer interaction} \hfill \\
\textit{Review comment usefulness -} Studies have investigated reviewer interaction through review comments. According to the primary studies, the usefulness of comments is determined by the changes caused by them~\citeS{rahman2017predicting,pangsakulyanont2014assessing}. For example, a useless comment can be an inquiry about the code that leads to no change. However, such discussions can lead to knowledge sharing. Therefore, the semantics of comments should also be considered when determining their usefulness. \\
\textit{Knowledge sharing -} Experienced reviewers possess tacit knowledge that is difficult to formalize and convey to novice programmers. Systems that could mine this knowledge from reviews would be an interesting avenue for research. It could be good if reviewers get a good mix of familiar and unknown code reviews to expand their expertise over time. Existing studies~\citeS{24, Rigby:2012, Czerwonka:2015, 2, 27, Kononenko:2016} show that the expertise of reviewers is codified in reviews, but making that expertise tangible and accessible is still an open question. 
In addition, a review comment can sometimes contain links that provide additional information that could contribute to knowledge sharing. \\
\textit{Human aspects -} When discussing reviewer interactions, human aspects received much attention in the reviewed primary studies. However, the investigation of review dynamics, social interactions, and review performance is focused on OSS projects. It is not known if such interactions differ in proprietary projects.
\begin{mybox}{Reviewer interaction}
    \begin{enumerate}
    \item[]\textbf{Review comment usefulness}
        \item[Q1] How could semantics to identify the usefulness of code review comments? 
        \item[] \textbf{Knowledge sharing}
            \item[Q2] How could the expertise of reviewers be mined through code review comments and made accessible to less experienced reviewers? 
            \item[Q3] How can links between actors and information objects created in MCR be used to understand the review process better or extract useful information about the developed product?
       \item[]\textbf{Human aspects}
        \item[Q4] To what extent is the impact of social interactions on review performance specific to project type (OSS/proprietary)?
    \end{enumerate}
    \end{mybox}

\subsubsection{Review decision} \hfill \\
We have identified studies that automate code reviews. When static code analysis tools are used for automation, the rules are visible and explicit (white box). In contrast, the decision rules are not easily interpretable when using deep learning approaches (black box). 
\begin{mybox}{Reviewer decision}
    \begin{enumerate}
    \item[] \textbf{Automation}
        \item[Q1] How can code review automation, based on deep learning approaches integrate explainability and transparency of accept/reject decisions?
    \end{enumerate}
    \end{mybox}
    
\subsubsection{Overall code review process} \hfill \\
The future research directions and research questions that could not be directly mapped to any specific step in the code review process are categorized as overall topics. \\
\textit{Human aspects -} The difference between core and irregular reviewers has been studied mostly in the context of open source (OSS)~\citeS{Baysal:2012b, 18, Kerzazi:2016, Bosu:2012, pinto2018gets}. In our survey, the respondents perceived the difference between core and irregular reviewers as unimportant. However, the survey respondents were mostly from companies working with proprietary projects. It would be interesting to investigate if human factors matter more in OSS than in proprietary software development. 
The survey respondents considered the research on human aspects not as relevant as the research on other themes. We pose that, since most of the research on organizations and human factors is performed in OSS projects, research on human aspects in companies mainly working with proprietary projects is still very relevant as the human factors might differ between OSS and proprietary contexts. \\
\textit{Innersourcing -} We have seen MCR research in OSS and in commercial projects. However, we do know very little about MCR in Innersourcing. Similar to the open source way of working, Innersourcing promotes contributions across different projects within an organization. A delay or incomplete review may discourage an Innersource contributor. Human factors may be more relevant for commercial projects in the context of Innersourcing. \\
\textit{Code reuse -} Common assets are code that is reused within an organization. It would be interesting to investigate how changes made to common assets should be reviewed. For example, should reviewers from teams that use the common assets be invited to review new changes?\\
\begin{mybox}{Overall}
    \begin{enumerate}
    \item[]\textbf{Human aspects}
        \item[Q1]  Are the OSS studies' findings not applicable to proprietary software development?
        \item[]\textbf{Innersourcing}
        \item[Q2] How does code review performance (review time, thoroughness, etc.) impact innersource contributions?
        \item[]\textbf{Common assets review}
        \item[Q3] How should changes in common assets be reviewed?
    \end{enumerate}
    \end{mybox}

\subsection{A comparison with other reviews and surveys}
\begin{figure}
    \centering
    \includegraphics[width=0.5\columnwidth]{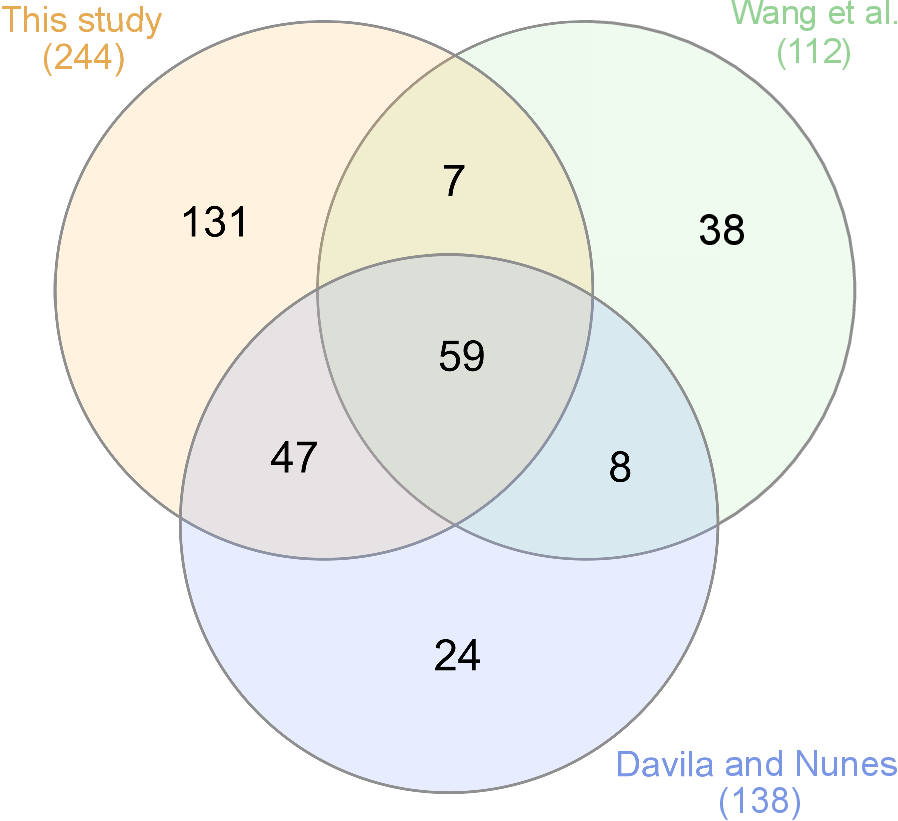}
    \caption {Unique and common primary studies in MCR reviews with wide scope}
    \label{fig:three_venn_newsearch}
\end{figure}

Figure~\ref{fig:three_venn_newsearch} illustrates the overlap and differences between the primary studies identified in this paper and the works by Wang et al., and Davila and Nunes. Interestingly, even though the search period and aims of the three studies are comparable in scope, the included primary studies are quite diverse. Wohlin et al.~\cite{wohlin2013reliability} have made similar observations when comparing the results of two comparable mapping studies.
While a detailed analysis, as done by Wohlin et al., is out of the scope of this paper, we speculate that the main reason for the divergence of primary studies is the emphasis on different keywords in the respective searches. While Wang et al., and Davila and Nunes included "inspection" in their search string, we explicitly did not. The term is associated with traditional code inspections, which is a different process than MCR, as explained in Section~\ref{sec:background}. Rather, we included terms that are associated with MCR, like "pull request" and "patch acceptance". Wang et al. excluded also papers that were not published in a high impact venue, likely leading to the lower number of included primary studies. We also explicitly did not exclude any papers based on their quality. Instead, we provide an assessment of the studies' rigor and relevance~\cite{ivarsson2011method}. 

As described in Section~\ref{Sec:Quality}, there is a lack of involvement of human subjects in code review research. As code review is a human-centric process, researchers should involve more human subjects to evaluate the feasibility of code review interventions. This research gap has also been observed by Davila and Nunes' review~\cite{davila2021systematic}, who called for more user studies on MCR. Moreover, researchers should consider more data sources in addition to analyzing repositories to achieve triangulation, strengthening the conclusions that can be drawn on the developed interventions. This is analogous to the observation by Davila and Nunes~\cite{davila2021systematic}, who found that a majority of MCR research focuses on particular open source projects (Qt, Openstack, and Android) or is conducted in large companies such as Microsoft and Google. 
MCR practices and motivations in companies working primarily with proprietary projects may be different than in open source projects. In companies, the contributions and reviews are not voluntary work; furthermore, it is more likely that the reviewers and contributors known each other. Moreover, there may be domain or organization-specific requirements that may not be present in open source projects. Therefore, there is a need to investigate more MCR in the context of proprietary software projects.
It is also worth mentioning that there is a lack of studies in small and medium-sized companies. Such new studies could shed light on the question if the knowledge accumulated through MCR is also present in small and medium companies' projects~\cite{davila2021systematic}. 

The surveyed practitioners were most positive about the impact of code reviews on product quality and human aspects (IOF) and modern code review process properties (CRP). Therefore, given the relatively low number of studies in these themes, we suggest conducting more research investigating how MCR affects and is affected. Davila and Nunes~\cite{davila2021systematic} share a similar insight, calling for more research on MCR process improvement.  

Previous studies that surveyed practitioners and analyzed the impact of software engineering research~\cite{lo2015practitioners,carver2016practitioners} could not establish any correlation (positive or negative) between research impact and practitioner interest. These observations are to a large extent in line with our results (Section~\ref{sec:compImpact}). There is only one statistical significant correlation between the Support Systems theme and research impact, indicating that statements containing publications with high citation count were considered as important to investigate (and vice-versa).
In Section~\ref{sec:practice}, we report also that the past surveys on practitioner perception found that 67-71\% of the research was seen positively. Looking at the results in Figure~\ref{fig:ratings}, we observe that 24 statements have more negative than positive ratings and 22 statements have more positive than negative ratings (48\% positive). This is considerably lower than in the previous surveys. However, it could be attributed to the difference in the administered survey instruments. In our survey, the participants \emph{had} to distribute negative, neutral, and positive perception according to a predefined distribution, i.e., they could not find all research negative or positive. 


\subsection{Validity threats}\label{Sec:vt}
Our philosophical worldview is pragmatic, which is often adopted by software engineering researchers~\cite{petersen2013worldviews}. We use therefore a mixed-method approach (systematic mapping study and survey) to answer our research questions. The commonly used categorizations to analyze and report validity threats, such as internal, external, conclusion and construct validity of the post-positivist worldview (such as Wohlin et al~\cite{wohlin_guidelines_2014}), are adequate for quantitative research. However, they either do not capture all relevant threats for qualitative research, or are formulated in a way that is not compatible with an interpretative worldview. The same argument can be made for threat categorizations originating from the interpretative worldview (such as Runeson et al.~\cite{runeson_guidelines_2009}). Petersen and Gencel~\cite{petersen2013worldviews} have therefore proposed a complementary validity threat categorization, based on work by Maxwell~\cite{maxwell1992understanding}, that is adequate for a pragmatic worldview. We structure the remainder of the section according to their categorization and discuss validity threats w.r.t. each research method. Please note that repeatability (reproducibility/dependability) of this research is a function of all the threat categories together~\cite{petersen2013worldviews}, and therefore not discussed individually.

\subsubsection{Descriptive validity}
Threats in this category concern the factual accuracy of the made observations~\cite{petersen2013worldviews}. We have designed and used a data extraction form to collect the information from the reviewed studies. We copied contribution statements that can be traced back to the original studies. Furthermore, we have piloted the survey instrument with three practitioners to identify functional defects and usability issues. Extraction form, survey instrument and collected data are available in an online repository~\cite{deepika_badampudi_2022_7066821}. 

\subsubsection{Theoretical validity}\label{sec:theorecticalv}
Threats in this category concern the ability to capture practically the intended theoretical concepts, which includes the control of confounding factors and biases~\cite{petersen2013worldviews}.

\textbf{Study identification.} During the search we could have missed papers that could have been relevant but were not identified by the search string. We addressed this threat with a careful selection of keywords and not limiting the scope of the search to a particular population, comparison or outcome (PICO criteria~\cite{kitchenham_guidelines_2007}). For the intervention criterion, we used variants of terms that we deemed relevant and associated with modern code reviews. Due to this association, we did not choose keywords for the population criterion (e.g. "software engineering") as they could have potentially reduced the number of relevant search hits. We have compared our primary studies with the set of other systematic literature studies (see Figure~\ref{fig:three_venn_newsearch}). We have identified 136 studies that the other reviews missed, while not identifying 69 studies that were found by the other reviews. Hence, there is a moderate threat that we missed relevant studies.

\textbf{Study selection.} Researcher bias could have led to the wrongful exclusion of relevant papers. We addressed this threat by including all three authors in the selection process who reviewed an independent set of studies. To align our selection criteria, we established objective inclusion and exclusion criteria, which we piloted and refined when we found divergences in our selection (see \emph{Selection} in Section~\ref{SubSec:SMS}). Furthermore, we adopted an inclusive selection process, postponing the decision on exclusion for unsure papers to the data extraction step, when it would be clear that the study did not contain the information we required to answer our research questions. When we excluded a paper, we documented the decision with the particular exclusion criterion. 

\textbf{Data extraction.} Researcher bias could have led to a incorrect extraction of data. All three authors were involved in the data extraction as well. We also conducted two pilot extractions to gain a common interpretation of the extraction items. We revised the description of rigor and relevance criteria based on the pilot process. After the pilot process we continued to extract data from primary studies with an overlap of 20\% where we achieved high consensus.

\textbf{Statement order.} All survey participants received the statements in the same order. The participants may have tended to agree more to statements listed at the beginning than at the end of the survey. Our survey instrument was designed in such a way that the participants could change their rating anytime, i.e. also when they have seen all statements. Looking at our results, the themes that were judged early in the survey (IOF) seem to have received more agreement than later themes (SS). However, participants have provided a consistent rating for the human factor related statements, independently of whether or not they appear in early or late positions in the survey. Therefore we assess the likelihood of this risk as low.

\subsubsection{Internal generalizability}
Threats in this category concern the degree to which inferences from the collected data (quantitative or qualitative) are reasonable~\cite{petersen2013worldviews}. 

\textbf{Statement ranking and factor analysis.} We followed the recommendations for conducting the Q-Methodology~\cite{zabala2016bootstrapping}, including factor analysis and interpretation. In addition, we report in detail how we interpret the quantitative results of the survey, providing a chain of evidence for our argumentation and conclusions. 

\textbf{Research amount and impact, and practitioners perceptions.} There is a risk that practitioners understood that they need to judge if the topic in a statement affects them, rather than whether or not research on the topic is important. We counteracted this threat by designing the survey instrument in a way that reminds the respondents what the purpose of the ranking of statements is. Furthermore, the free-text answers in the survey provide a good indication that the respondents correctly understood the ranking task.

\textbf{Identification of research roadmap.} We have based our analysis on the contributions reported by the original authors of the studies. In contrast, the gaps we highlighted are based on what has \emph{not} been reported (i.e. researched). As such, the proposed research agenda contain speculation on what might be fruitful to research. However, we do provide argumentation and references to the original studies, allowing readers to follow our reasoning.

\subsubsection{External generalizability}
Threats in this category concern the degree to which inferences from the collected data (quantitative or qualitative) can be extended to the population from which the sample was drawn. 

\textbf{Definition of statements.} There is a risk that the statements were formulated either too generic or too specific, not reflecting all aspects of the research studies they represent. The statements were defined based on the primary studies collected in the 2018 SMS. We then extended the SMS to include papers published until 2021, classifying all new studies, except one, under the existing statements, which indicates that the initially defined statements were still useful after four years of research. It is however likely that with the advancement of MCR practice, new research emerges that requires also an update of the statements if they survey is replicated in the future. 

\textbf{Survey sample.} As discussed in Section~\ref{sec:background}, the main difference between proprietary and open source software development, in relation to MCR, is the purpose of the MCR practice. The respondents of our survey mainly work in companies that primarily work in proprietary projects. In this context, the main purpose of MCR is knowledge dissemination rather than building relationships to core developers~\citeS{bosu2016process}. Indeed, we observe in our survey results that research aspects of human factors in relation to MCR are perceived as less important (see Section~\ref{subsec:agreementlevel}). However, the factor analysis in Section~\ref{Sec:differentViews} provides a more differentiated view, based on the profile of the survey respondents. For example, senior roles are more favourable towards human factors research in MCR than respondents with less professional experience. Nevertheless, future work could replicate the survey in open source communities, allowing a differential analysis.

\textbf{Identification of research roadmap.} While the inferences we draw from the reviewed studies may be sound \emph{within} the sample we studied, there is a moderate threat that the future research we propose has already been conducted in the studies we did not review (see discussion on study identification in Section~\ref{sec:theorecticalv}).

\subsubsection{Interpretive validity}
Threats in this category concern the validity of the conclusions that are drawn by the researchers by interpreting the data~\cite{petersen2013worldviews}.

\textbf{Definition of themes.} Researcher bias could have lead to an inaccurate classification of the studies in the SMS. We divided the primary studies for analysis among the authors. Each paper that was analysed by one author was reviewed by the two other authors, and disagreements were discussed until a change in the classification was made or consensus was reached.

\textbf{Definition of statements.} Also the formulation of statements representing a study, used in the survey, could have been affected by researcher bias. To address this treat, we followed an iterative process in which we revised the statements and the association of papers to these statements. All three authors were involved in this process and reviewed each others formulations and classifications to check for consistency as well as allow for different perspectives on the material. There are two statements where more than one aspect is introduced: in the HOF theme, "age and experience" and "performance and quality" in the SS theme. A respondent may have an opinion on just one of the aspects and therefore misrepresent the rating of the other aspect. We asked the practitioners to provide explanations for extreme ratings i.e., -3 and +3 which makes it possible to know which aspects the practitioners focused on. However, such explanations are not available for ratings other than -3 to +3. Since only two out of 46 statements are affected we judge the risk of misrepresentation as low.\\

\textbf{Identification of research roadmap.} Finally, the identification of gaps in the MCR research corpus could have been affected by researcher bias. The first and second author conducted a workshop in which they independently read the MCR contributions in Section~\ref{sec:mcrcontributions}. Then, they discussed their ideas of what questions have \emph{not} been answered by the reviewed research and which questions would be interesting to find an answer for, especially if there is some support from the survey results that a particular statement was perceived important to investigate by the practitioners. This initial formulation of research gaps was reviewed by the third author.

\section{Conclusions} \label{sec:conclusions}
In this paper, we conducted a systematic mapping study and a survey to provide an overview of the different research themes on Modern Code Reviews (MCR) and analyze the practitioners’ opinions on the importance of those themes. Based on the juxtaposition of these two perspectives on MCR research, we outline an agenda for future research on MCR that is based on the identified research gaps and the perceived importance by practitioners.

We have extracted the research contributions from 244 primary studies and
summarized 15 years of MCR research in evidence briefings that can contribute to the knowledge transfer from academic research to practitioners. The five main themes of MCR research are: (1) support systems for code reviews (SS), (2) impact of code reviews on product quality and human aspects (IOF), (3) modern code review process properties (CRP), (4) impact of software development processes, patch characteristics, and tools on modern code reviews (ION), and (5) human and organizational factors (HOF).

We conducted a survey to collect practitioners' opinions about 46 statements representing the research in the identified themes. As a result, we learned that practitioners are most positive about the CRP and IOF theme, with special focus on the impact of code reviews on product quality. However, these themes  represent a minority of the reviewed MCR research (66 primary studies). At the same time, the respondents are most negative about human factor- (HOF) and support systems-related (SS) research, which constitute together a majority of the reviewed research (108 primary studies). These results indicate that there is a misalignment between the state-of-the-art and the themes deemed important by most respondents of our survey. In addition, we found that the research that has been perceived positively by practitioners is generally also more frequently cited, i.e., has a larger research impact.

Finally, as there has been an increased interest in reviewing MCR research in recent years, we analyzed other systematic literature reviews and mapping studies. Due to the different research questions of the respective studies, there is a varying overlap of the reviewed primary studies. Still, we find our observations on the potential gaps in MCR research corroborated. Analyzing the data extracted from the reviewed primary studies and guided by the answers from the survey, we propose nineteen new research questions we deem worth investigating in future MCR research.

\section{Acknowledgments} We would like to acknowledge that this work was supported by the Knowledge Foundation through the projects SERT – Software
Engineering ReThought and OSIR Open-source inspired reuse (reference number 20190081) at Blekinge Institute of Technology, Sweden. We would also like to acknowledge all practitioners who contributed to our investigation.

\bibliographystyle{ACM-Reference-Format}
\bibliography{references}



\bibliographystyleS{ACM-Reference-Format}
\bibliographyS{references}

\end{document}